# Adaptive Quadrature Detection for Multicarrier Continuous-Variable Quantum Key Distribution

Laszlo Gyongyosi

[1] Quantum Technologies Laboratory, Department of Telecommunications
*Budapest University of Technology and Economics*
2 Magyar tudosok krt, Budapest, *H*-1117, Hungary
[2] MTA-BME Information Systems Research Group
*Hungarian Academy of Sciences*
7 Nador st., Budapest, *H*-1051, Hungary

gyongyosi@hit.bme.hu

**Abstract**

We propose the adaptive quadrature detection for multicarrier continuous-variable quantum key distribution (CVQKD). A multicarrier CVQKD scheme uses Gaussian subcarrier continuous variables for the information conveying and Gaussian sub-channels for the transmission. The proposed multicarrier detection scheme dynamically adapts to the sub-channel conditions using a corresponding statistics which is provided by our sophisticated sub-channel estimation procedure. The sub-channel estimation phase determines the transmittance coefficients of the sub-channels, which information are used further in the adaptive quadrature decoding process. We define the technique called subcarrier spreading to estimate the transmittance conditions of the sub-channels with a theoretical error-minimum in the presence of a Gaussian noise. We introduce the terms of single and collective adaptive quadrature detection. We also extend the results for a multiuser multicarrier CVQKD scenario. We prove the achievable error probabilities, the signal-to-noise ratios, and quantify the attributes of the framework. The adaptive detection scheme allows to utilize the extra resources of multicarrier CVQKD and to maximize the amount of transmittable valuable information in diverse measurement and transmission conditions. The framework is particularly convenient for experimental CVQKD scenarios.

# 1 Introduction

The continuous-variable quantum key distribution (CVQKD) protocols allow for legal parties to establish an unconditionally secure communication through standard telecommunication devices and networks [1-13], 16-20]. In comparison with discrete-variable (DV) QKD protocols, the encoding and decoding processes do not require single-photon devices and can be implemented through standard devices of optical telecommunications [14], [21-22], [26-33]. In a CVQKD protocol, the information is generally conveyed through the position and momentum quadratures of quantum states. These quadratures together identify a continuous variable [CV] state in the phase space (referred to as *single-carrier* CV). In a practical CVQKD modulation, the quadratures have a Gaussian random distribution, and the presence of the eavesdropper adds a white Gaussian noise to the quadrature transmission [16-20]. At the receiver, the noisy quadratures are detected by a homodyne (e.g., a position or a momentum quadrature is measured) or a heterodyne (e.g., both position and momentum quadrates are measured) measurement apparatus. Although the practical implementation of CVQKD requires no specialized devices, the performance of CVQKD protocols still brings up several challenges and demands significant improvements. Because the single-carrier CVQKD transmission allows no to exploit several invaluable essential resources in the transmission phase, the multicarrier CVQKD has been proposed [2-8]. In comparison with single-carrier CVQKD, the multicarrier transmission uses Gaussian subcarrier CVs for the information conveying, which are sent through Gaussian sub-channels. Particularly, the Gaussian sub-channels are derived from the physical Gaussian link via a CV unitary [2]. The multicarrier CVQKD modulation has been introduced via the adaptive multicarrier quadrature division (AMQD) framework [2], which also has been extended to a multiuser multicarrier CVQKD setting through the multiuser quadrature allocation scheme [3]. The multicarrier transmission possesses several benefits over single-carrier transmission, such as improved noise tolerance, higher secret key rates, enhanced security thresholds, and extended transmission distances [2-8]. The multicarrier modulation also allows to exploit those additional degrees of freedom of the transmission that are not available in a single-carrier scenario. Specifically, these extra resources allow us to define several new phenomena that cannot be utilized in a single-carrier setting, such as singular layer transmission [4], improved security thresholds [5], multidimensional manifold extraction [6], and subcarrier domain achievement [7]. These additional resources are provided by the extra degree of freedom of the multicarrier CVQKD scheme. In this work, we are focusing on the decoding mechanism and provide an efficient decoding framework for multicarrier CVQKD that allows for the legal parties to utilize and manifest the potential of the multicarrier modulation.

We define the *adaptive quadrature detection* for multicarrier CVQKD. The adaptive detection dynamically adapts to the channel conditions using a corresponding statistics of the sub-channels. This statistics is provided for the adaptive quadrature decoding process by our sophisticated *sub-channel estimation* procedure. Precisely, the proposed estimation determines the transmittance coefficients of the sub-channels, which are used further in the process of adaptive quadrature decoding. In particular, the conditions of the sub-channels are determined by *pilot-subcarrier* CV quantum states, which carry no valuable information and used only in a dedicated calibration phase prior to the private information transmission of the protocol run. We introduce the technique called *subcarrier spreading*, which uses a subcarrier flow to estimate the transmittance conditions of the sub-channels with a theoretical minimum error probability. The adaptive quadrature detection procedure can be applied for single or collective, homodyne, or heterodyne measurement settings, receptively. We also extend the adaptive quadrature detection for a multiuser multicarrier CVQKD scenario. We derive the details of the adaptive decoding procedure, prove the achievable error probabilities and signal-to-noise ratios (SNRs), and quantify the attributes of the quadrature detection scheme. The adaptive detection scheme allows to perform the decoding of the quadratures with maximal efficiency at diverse channel conditions, and for arbitrarily distributed channel coefficients. In particular, it also allows to utilize the extra resources of multicarrier CVQKD and to maximize the amount of transmittable valuable information in diverse measurement and transmission conditions for any multicarrier CVQKD scenario.

This paper is organized as follows. Section 2 summarizes some preliminary findings. Section 3 proposes sub-channel estimation procedure for multicarrier CVQKD. Section 4 discusses the adaptive quadrature detection scheme. Section 5 extends the adaptive quadrature detection for a multiuser multicarrier scenario. Finally, Section 6 concludes the results. Supplemental information is included in the Appendix.

## 2 Preliminaries

In Section 2, we briefly summarize the notations and basic terms. For further information, see the detailed descriptions of [2–7].

### 2.1 Basic Terms and Definitions

#### 2.1.1 Multicarrier CVQKD

In this section we very briefly summarize the basic notations of AMQD from [2]. The following description assumes a single user, and the use of $n$ Gaussian sub-channels $\mathcal{N}_i$ for the transmission of the subcarriers, from which only $l$ sub-channels will carry valuable information.

In the single-carrier modulation scheme, the $j$-th input single-carrier state $|\varphi_j\rangle = |x_j + \mathrm{i}p_j\rangle$ is a Gaussian state in the phase space $\mathcal{S}$, with i.i.d. Gaussian random position and momentum quadratures $x_j \in \mathbb{N}\left(0, \sigma_{\omega_0}^2\right)$, $p_j \in \mathbb{N}\left(0, \sigma_{\omega_0}^2\right)$, where $\sigma_{\omega_0}^2$ is the modulation variance of the quadratures. In the multicarrier scenario, the information is carried by Gaussian subcarrier CVs, $|\phi_i\rangle = |x_i + \mathrm{i}p_i\rangle$, $x_i \in \mathbb{N}\left(0, \sigma_{\omega}^2\right)$, $p_i \in \mathbb{N}\left(0, \sigma_{\omega}^2\right)$, where $\sigma_{\omega}^2$ is the modulation variance of the subcarrier quadratures, which are transmitted through a noisy Gaussian sub-channel $\mathcal{N}_i$. Precisely, each $\mathcal{N}_i$ Gaussian sub-channel is dedicated for the transmission of one Gaussian subcarrier CV from the $n$ subcarrier CVs. (*Note*: index $i$ refers to a subcarrier CV, index $j$ to a single-carrier CV, respectively.)

The single-carrier state $|\varphi_j\rangle$ in the phase space $\mathcal{S}$ can be modeled as a zero-mean, circular symmetric complex Gaussian random variable $z_j \in \mathcal{CN}\left(0, \sigma^2_{\omega_{z_j}}\right)$, with a variance

$$\sigma^2_{\omega_{z_j}} = \mathbb{E}\left[|z_j|^2\right] = 2\sigma^2_{\omega_0}, \tag{1}$$

and with i.i.d. real and imaginary zero-mean Gaussian random components

$$\mathrm{Re}(z_j) \in \mathbb{N}\left(0, \sigma^2_{\omega_0}\right), \; \mathrm{Im}(z_j) \in \mathbb{N}\left(0, \sigma^2_{\omega_0}\right). \tag{2}$$

In the multicarrier CVQKD scenario, let $n$ be the number of Alice's input single-carrier Gaussian states. Precisely, the $n$ input coherent states are modeled by an $n$-dimensional, zero-mean, circular symmetric complex random Gaussian vector

$$\mathbf{z} = \mathbf{x} + \mathrm{i}\mathbf{p} = (z_0, \ldots, z_{n-1})^T \in \mathcal{CN}(0, \mathbf{K_z}), \tag{3}$$

where each $z_j$ is a zero-mean, circular symmetric complex Gaussian random variable

$$z_j \in \mathcal{CN}\left(0, \sigma^2_{\omega_{z_j}}\right), \; z_j = x_j + \mathrm{i}p_j. \tag{4}$$

In the first step of AMQD, Alice applies the inverse FFT (fast Fourier transform) operation to vector $\mathbf{z}$ (see (3)), which results in an $n$-dimensional zero-mean, circular symmetric complex Gaussian random vector $\mathbf{d}$, $\mathbf{d} \in \mathcal{CN}(0, \mathbf{K_d})$, $\mathbf{d} = (d_0, \ldots, d_{n-1})^T$, precisely as

$$\mathbf{d} = F^{-1}(\mathbf{z}) = e^{\frac{\mathbf{d}^T \mathbf{A}\mathbf{A}^T \mathbf{d}}{2}} = e^{\frac{\sigma^2_{\omega_0}\left(d_0^2 + \ldots + d_{n-1}^2\right)}{2}}, \tag{5}$$

where

$$d_i = x_{d_i} + \mathrm{i}p_{d_i}, \; d_i \in \mathcal{CN}\left(0, \sigma^2_{d_i}\right), \tag{6}$$

where $\sigma^2_{\omega_{d_i}} = \mathbb{E}\left[|d_i|^2\right] = 2\sigma^2_\omega$, thus the position and momentum quadratures of $|\phi_i\rangle$ are i.i.d. Gaussian random variables with a constant variance $\sigma^2_\omega$ for all $\mathcal{N}_i, i = 0, \ldots, l-1$ sub-channels:

$$\mathrm{Re}(d_i) = x_{d_i} \in \mathbb{N}\left(0, \sigma^2_\omega\right), \; \mathrm{Im}(d_i) = p_{d_i} \in \mathbb{N}\left(0, \sigma^2_\omega\right), \tag{7}$$

where $\mathbf{K_d} = \mathbb{E}\left[\mathbf{dd}^\dagger\right]$, $\mathbb{E}[\mathbf{d}] = \mathbb{E}\left[e^{\mathrm{i}\gamma}\mathbf{d}\right] = \mathbb{E}e^{\mathrm{i}\gamma}[\mathbf{d}]$, and $\mathbb{E}\left[\mathbf{dd}^T\right] = \mathbb{E}\left[e^{\mathrm{i}\gamma}\mathbf{d}\left(e^{\mathrm{i}\gamma}\mathbf{d}\right)^T\right] = \mathbb{E}e^{\mathrm{i}2\gamma}\left[\mathbf{dd}^T\right]$ for any $\gamma \in [0, 2\pi]$. The $\mathbf{T}(\mathcal{N})$ transmittance vector of $\mathcal{N}$ in the multicarrier transmission is

$$\mathbf{T}(\mathcal{N}) = [T_0(\mathcal{N}_0), \ldots, T_{n-1}(\mathcal{N}_{n-1})]^T \in \mathcal{C}^n, \tag{8}$$

where

$$T_i(\mathcal{N}_i) = \mathrm{Re}(T_i(\mathcal{N}_i)) + \mathrm{i}\,\mathrm{Im}(T_i(\mathcal{N}_i)) \in \mathcal{C}, \tag{9}$$

is a complex variable, which quantifies the position and momentum quadrature transmission (i.e., gain) of the $i$-th Gaussian sub-channel $\mathcal{N}_i$, in the phase space $\mathcal{S}$, with real and imaginary parts

$$0 \leq \operatorname{Re} T_i(\mathcal{N}_i) \leq 1/\sqrt{2}, \text{ and } 0 \leq \operatorname{Im} T_i(\mathcal{N}_i) \leq 1/\sqrt{2}. \tag{10}$$

Particularly, the $T_i(\mathcal{N}_i)$ variable has the squared magnitude of

$$|T_i(\mathcal{N}_i)|^2 = \operatorname{Re} T_i(\mathcal{N}_i)^2 + \operatorname{Im} T_i(\mathcal{N}_i)^2 \in \mathbb{R}, \tag{11}$$

where

$$\operatorname{Re} T_i(\mathcal{N}_i) = \operatorname{Im} T_i(\mathcal{N}_i). \tag{12}$$

The Fourier-transformed transmittance of the $i$-th sub-channel $\mathcal{N}_i$ (resulted from CVQFT operation at Bob) is denoted by

$$|F(T_i(\mathcal{N}_i))|^2. \tag{13}$$

The $n$-dimensional zero-mean, circular symmetric complex Gaussian noise vector $\Delta \in \mathcal{CN}(0, \sigma_\Delta^2)_n$, of the quantum channel $\mathcal{N}$, is evaluated as

$$\Delta = (\Delta_0, \ldots, \Delta_{n-1})^T \in \mathcal{CN}(0, \mathbf{K}_\Delta), \tag{14}$$

where

$$\mathbf{K}_\Delta = \mathbb{E}[\Delta \Delta^\dagger], \tag{15}$$

with independent, zero-mean Gaussian random components

$$\Delta_{x_i} \in \mathbb{N}(0, \sigma_{\mathcal{N}_i}^2), \text{ and } \Delta_{p_i} \in \mathbb{N}(0, \sigma_{\mathcal{N}_i}^2), \tag{16}$$

with variance $\sigma_{\mathcal{N}_i}^2$, for each $\Delta_i$ of a Gaussian sub-channel $\mathcal{N}_i$, which identifies the Gaussian noise of the $i$-th sub-channel $\mathcal{N}_i$ on the quadrature components $x_i, p_i$ in the phase space $\mathcal{S}$. Thus $F(\Delta) \in \mathcal{CN}(0, \sigma_{\Delta_i}^2)$, where

$$\sigma_{\Delta_i}^2 = 2\sigma_{\mathcal{N}_i}^2. \tag{17}$$

The CVQFT-transformed noise vector can be rewritten as

$$F(\Delta) = (F(\Delta_0), \ldots, F(\Delta_{n-1}))^T, \tag{18}$$

with independent components $F(\Delta_{x_i}) \in \mathbb{N}(0, \sigma_{\mathcal{N}_i}^2)$ and $F(\Delta_{p_i}) \in \mathbb{N}(0, \sigma_{\mathcal{N}_i}^2)$ on the quadratures, for each $F(\Delta_i)$. Precisely, it also defines an $n$-dimensional zero-mean, circular symmetric complex Gaussian random vector $F(\Delta) \in \mathcal{CN}(0, \mathbf{K}_{F(\Delta)})$ with a covariance matrix

$$\mathbf{K}_{F(\Delta)} = \mathbb{E}[F(\Delta) F(\Delta)^\dagger]. \tag{19}$$

## 3 Sub-channel Estimation for Multicarrier CVQKD

In the first part, we study the process of building multicarrier channel statistics. In the second part, we introduce the subcarrier spreading transmission technique, which minimizes the error probability of the estimation process in the presence of a Gaussian noise.

## 3.1 Sub-channel and Single-Carrier Channel Model

**Theorem 1** (*Sub-channel estimation of multicarrier CVQKD*). *For any $p_i \in \mathcal{C}$, $|p_i| > 0$ pilot-subcarrier CV, the $\mathcal{S}_i(\mathcal{N}_i) \in \mathcal{C}$ sufficient statistic for the estimation of $F(T_i(\mathcal{N}_i))$ of $\mathcal{N}_i$ is $\mathcal{S}_i(\mathcal{N}_i) = \varsigma_i^{\dagger} p_i' = F(T_i(\mathcal{N}_i)) + F'(\Delta)$, where $\varsigma_i = p_i / |p_i|^2$, $p_i' = F(T_i(\mathcal{N}_i))p_i + F(\Delta)$, $F(\Delta) \in \mathcal{CN}(0, 2\sigma_{\mathcal{N}}^2)$, $F'(\Delta) \in \mathcal{CN}(0, 2\sigma_{\mathcal{N}}^2 / |p_i|^2)$.*

*The $\mathcal{S}(\mathcal{N}_j) \in \mathcal{C}$ of $\mathcal{N}_j$ is $\mathcal{S}(\mathcal{N}_j) = \varsigma_j^{\dagger} \mathbf{q}_j' = A_j + F'(\Delta)$, where $\mathbf{q}_j = (p_{j,0}, \ldots, p_{j,l-1})^T \in \mathcal{C}^l$, $\mathbf{q}_j' = (p_{j,0}', \ldots, p_{j,l-1}')^T \in \mathcal{C}^l$ are pilot-subcarrier CV vectors, $A_j = \frac{1}{l} \left( \sum_{i=0}^{l-1} F(T_{j,i}(\mathcal{N}_{j,i})) \right) \in \mathcal{C}$, $\varsigma_j = \mathbf{q}_j / |\mathbf{q}_j|^2$, and $F'(\Delta) \in \mathcal{CN}(0, 2\sigma_{\mathcal{N}}^2 / |\mathbf{q}_j|^2)$.*

*Proof.*
The proof assumes the use of $n$ sub-channels from which $l$, $l < n$ sub-channels are selected via the proposed estimation process. For the sub-channel selection criteria, see the properties of AMQD modulation [2]. Throughout the manuscript we focus only on the $l$ sub-channels that are used for valuable private information transmission in the multicarrier CVQKD protocol run. The $i$-th sub-channel is referred via $\mathcal{N}_i$. From the $l$ sub-channels, a $j$-th single-carrier channel is derived, depicted via $\mathcal{N}_j$ [2-7]. In general, the referred channel output variables and vectors are yielding from a $M_{\text{hom}}$ homodyne or $M_{\text{het}}$ heterodyne measurement apparatus, respectively. These quantities are then post-processed via the corresponding functions and operators.

Let $p_i \in \mathcal{C}$ be a complex scalar variable known by Alice and Bob (also referred to as a *pilot-subcarrier CV*), which is used for the estimation of the $i$-th Gaussian sub-channel $\mathcal{N}_i$. (*Note*: the pilot CVs carry no valuable information and used in the sub-channel estimation phase to determine the channel transmittance coefficients).

In particular, in an AMQD multicarrier CVQKD setting [2], $p_i$ is defined via the inverse Fourier operation as

$$p_i = F^{-1}(p_j) = x_{p_i} + \mathrm{i} p_{p_i}, \tag{20}$$

which identifies a pilot-subcarrier CV with position and momentum quadratures $x_{p_i}$ and $p_{p_i}$, respectively, $\sigma_{\omega_{p_i}}^2 = \mathbb{E}\left[|p_i|^2\right] = 2\sigma_{\omega}^2$, for the estimation of the $F(T_i(\mathcal{N}_i))$ transmittance coefficient of the sub-channel $\mathcal{N}_i$, and where $p_j$ is referred to as the single-carrier pilot CV with quadrature components $x_{p_j}$ and $p_{p_j}$, evaluated as

$$p_j = F(p_i) = x_{p_j} + \mathrm{i} p_{p_j}, \tag{21}$$

where $F(\cdot)$ stands for the Fourier operator, $\sigma_{\omega_{p_j}}^2 = \mathbb{E}\left[|p_j|^2\right] = 2\sigma_{\omega_0}^2$.

The $p'_i$ output of $\mathcal{N}_i$ is expressed as
$$p'_i = F\left(T_i\left(\mathcal{N}_i\right)\right) p_i + F\left(\Delta\right). \tag{22}$$
Then let
$$\varsigma_i = \frac{p_i}{|p_i|^2}, \tag{23}$$
from which the $\mathcal{S}_i\left(\mathcal{N}_i\right)$ sufficient statistics [23-25] is defined as follows:
$$\begin{aligned}\mathcal{S}_i\left(\mathcal{N}_i\right) &= \varsigma_i^\dagger p'_i \\ &= \frac{p_i^\dagger}{|p_i|^2} p'_i \\ &= F\left(T_i\left(\mathcal{N}_i\right)\right) + F'\left(\Delta\right),\end{aligned} \tag{24}$$
where
$$F'\left(\Delta\right) \in \mathcal{CN}\left(0, 2\sigma_\mathcal{N}^2 / |p_i|^2\right). \tag{25}$$
To estimate $F\left(T_i\left(\mathcal{N}_i\right)\right)$ from $\mathcal{S}_i\left(\mathcal{N}_i\right)$, one can utilize the $\mathcal{E}$ mean square error [23-25] as
$$\mathcal{E} = \mathbb{E}\left[\left|F\left(T_i\left(\mathcal{N}_i\right)\right) - \zeta\left(F\left(T_i\left(\mathcal{N}_i\right)\right)\right)\right|^2\right], \tag{26}$$
where $\zeta\left(F\left(T_i\left(\mathcal{N}_i\right)\right)\right)$ is the linear estimate of $F\left(T_i\left(\mathcal{N}_i\right)\right)$, evaluated as
$$\zeta\left(F\left(T_i\left(\mathcal{N}_i\right)\right)\right) = \mathbb{E}\left[F\left(T_i\left(\mathcal{N}_i\right)\right) \big| \mathcal{S}_i\left(\mathcal{N}_i\right)\right]. \tag{27}$$
Without loss of generality, the minimum mean squared error $E$ is expressed as
$$E = \min\left(\mathcal{E}\right) = \frac{\mathbb{E}\left[\left|F\left(T_i(\mathcal{N}_i)\right)\right|^2\right] 2\sigma_\mathcal{N}^2}{\mathbb{E}\left[\left|F\left(T_i(\mathcal{N}_i)\right)\right|^2\right]|p_i|^2 + 2\sigma_\mathcal{N}^2}. \tag{28}$$
In particular, from the orthogonality property [23] follows that
$$\mathbb{E}\left[\left(\zeta\left(F\left(T_i\left(\mathcal{N}_i\right)\right)\right) - F\left(T_i\left(\mathcal{N}_i\right)\right)\right) \mathcal{S}_i\left(\mathcal{N}_i\right)^\dagger\right] = 0. \tag{29}$$
The $\zeta\left(F\left(T_i\left(\mathcal{N}_i\right)\right)\right)$ in (27) can be rewritten as
$$\zeta\left(F\left(T_i\left(\mathcal{N}_i\right)\right)\right) = C^\dagger \mathcal{S}_i\left(\mathcal{N}_i\right), \tag{30}$$
where $C \in \mathcal{C}$ refers to a constant complex variable, defined as
$$C = \frac{\mathbb{E}\left[\left|F\left(T_i(\mathcal{N}_i)\right)\right|^2\right]}{\mathbb{E}\left[\left|F\left(T_i(\mathcal{N}_i)\right)\right|^2\right] + 2\sigma_\mathcal{N}^2}. \tag{31}$$
Thus, $\zeta\left(F\left(T_i\left(\mathcal{N}_i\right)\right)\right)$ can be expressed as
$$\zeta\left(F\left(T_i\left(\mathcal{N}_i\right)\right)\right) = \left(\frac{\mathbb{E}\left[\left|F\left(T_i(\mathcal{N}_i)\right)\right|^2\right]}{\mathbb{E}\left[\left|F\left(T_i(\mathcal{N}_i)\right)\right|^2\right] + 2\sigma_\mathcal{N}^2}\right)^\dagger \left(F\left(T_i\left(\mathcal{N}_i\right)\right) + F'\left(\Delta\right)\right). \tag{32}$$
As follows, $\mathcal{S}_i\left(\mathcal{N}_i\right)$ is provably a sufficient statistics to achieve the linear estimation of $F\left(T_i\left(\mathcal{N}_i\right)\right)$ with minimum mean squared error $E$.

The results can be extended to the transmission of the $j$-th single-carrier Gaussian CV as follows.

The single-carrier level channel $\mathcal{N}_j$ utilizes $l$ Gaussian sub-channels, $\mathcal{N}_j = \left(\mathcal{N}_{j,0},...,\mathcal{N}_{j,l-1}\right)^T$, and the output $p'_j$ is evaluated as

$$p'_j = A_j p_j + F(\Delta), \qquad (33)$$

where $A_j$ is the transmittance coefficient of $\mathcal{N}_j$, averaged from the $l$ sub-channels,

$$A_j = \tfrac{1}{l}\left(\sum_{i=0}^{l-1} F\left(T_{j,i}\left(\mathcal{N}_{j,i}\right)\right)\right) \in \mathcal{C}, \qquad (34)$$

while $p_j \in \mathcal{CN}\left(0, 2\sigma^2_{\omega_0}\right)$ is the $j$-th Gaussian single-carrier input CV, and $p'_j \in \mathcal{CN}\left(0, 2\left(\sigma^2_{\omega_0} + \sigma^2_{\mathcal{N}}\right)\right)$, where $\sigma^2_{\mathcal{N}}$ is the separated quadrature noise variance of $\mathcal{N}_j$.

A $\mathbf{q}_j$ pilot-subcarrier CV vector is defined as

$$\mathbf{q}_j = \left(p_{j,0},...,p_{j,l-1}\right)^T \in \mathcal{C}^l, \qquad (35)$$

where $p_{j,i}$ is transmitted through the Gaussian sub-channel $\mathcal{N}_{j,i}$.

The $\mathbf{q}'_j$ output vector is as

$$\mathbf{q}'_j = A_j \mathbf{q}_j + F\left(\mathbf{v}(\Delta)\right) = \left(p'_{j,0},...,p'_{j,l-1}\right)^T \in \mathcal{C}^l, \qquad (36)$$

where $F\left(\mathbf{v}(\Delta)\right)$ is the $l$-dimensional noise vector

$$F\left(\mathbf{v}(\Delta)\right) = \left(F(\Delta_0),...,F(\Delta_{l-1})\right)^T \in \mathcal{CN}\left(0, 2\sigma^2_{\mathcal{N}} I\right), \qquad (37)$$

where $I$ is the $l \times l$ identity. The $\mathcal{S}(\mathcal{N}_j) \in \mathcal{C}$ sufficient statistics for $\mathcal{N}_j$ is then evaluated as

$$\begin{aligned}
\mathcal{S}(\mathcal{N}_j) &= \varsigma_j^\dagger \mathbf{q}'_j \\
&= \frac{\mathbf{q}_j^\dagger}{|\mathbf{q}_j|^2} \mathbf{q}'_j \\
&= A_j + F'(\Delta) \\
&= \tfrac{1}{l}\left(\sum_{i=0}^{l-1} F\left(T_{j,i}(\mathcal{N}_{j,i})\right)\right) + F'(\Delta),
\end{aligned} \qquad (38)$$

where

$$\varsigma_j = \frac{\mathbf{q}_j}{|\mathbf{q}_j|^2}, \qquad (39)$$

$$F'(\Delta) \in \mathcal{CN}\left(0, 2\sigma^2_{\mathcal{N}}/|\mathbf{q}_j|^2\right). \qquad (40)$$

Then similar to (31), $\mathbf{C} \in \mathcal{C}^l$ can be defined specifically as

$$\mathbf{C} = \frac{\mathbb{E}\left[|A_j|^2\right]}{\mathbb{E}\left[|A_j|^2\right]|\mathbf{q}_j|^2 + 2\sigma^2_{\mathcal{N}}} \mathbf{q}_j, \qquad (41)$$

from which $\zeta\left(A_j^2\right)$ of $A_j$ is yielded as

$$\zeta(A_j) = \mathbf{C}\mathcal{S}(\mathcal{N}_j)$$
$$= \frac{\mathbb{E}\left[|A_j|^2\right]}{\mathbb{E}\left[|A_j|^2\right]|\mathbf{q}_j|^2 + 2\sigma_\mathcal{N}^2} \mathbf{q}_j \frac{\mathbf{q}_j^\dagger}{|\mathbf{q}_j|^2} \mathbf{q}_j' \quad (42)$$
$$= \frac{\mathbb{E}\left[|A_j|^2\right]}{\mathbb{E}\left[|A_j|^2\right]|\mathbf{q}_j|^2 + 2\sigma_\mathcal{N}^2} \mathbf{q}_j'.$$

Thus, the $E$ minimum mean squared error is precisely

$$E = \frac{\mathbb{E}\left[|A_j|^2\right] 2\sigma_\mathcal{N}^2}{\mathbb{E}\left[|A_j|^2\right]|\mathbf{q}_j|^2 + 2\sigma_\mathcal{N}^2}. \quad (43)$$

In particular, the determination of $\mathcal{S}(\mathcal{N}_j)$ in (38) can be interpreted via a projection $\mathcal{P}$ in $\mathcal{C}^l$. The noisy vector $\mathbf{q}_j'$ is projected onto $\varsigma_i = \mathbf{q}_j/|\mathbf{q}_j|^2$, which results the scalar quantity $\mathcal{S}(\mathcal{N}_j) = \varsigma_j^\dagger \mathbf{q}_j' = A_j + F'(\Delta)$.

The projection of $\mathbf{q}_j'$ onto $\varsigma_i$ in the phase space $\mathcal{S}$ is illustrated in Fig. 1.

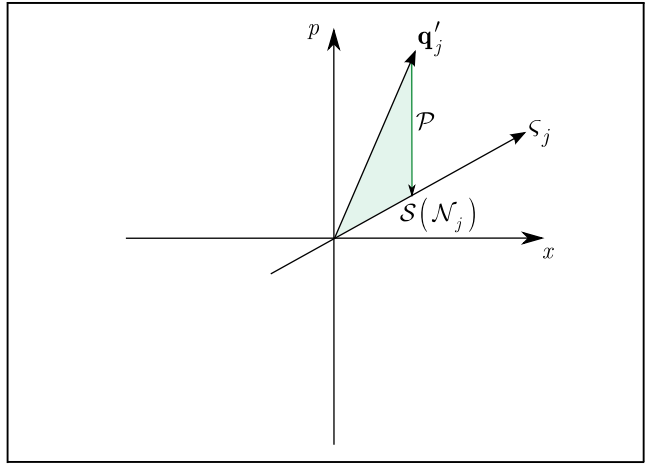

**Figure 1**. Determination of $\mathcal{S}(\mathcal{N}_j)$ sufficient statistics via projection $\mathcal{P}$ in $\mathcal{C}^l$. The noisy vector $\mathbf{q}_j'$ is projected onto $\varsigma_i$, which results the scalar quantity $\mathcal{S}(\mathcal{N}_j) = \varsigma_j^\dagger \mathbf{q}_j'$ ($x$, position quadrature; $p$, momentum quadrature).

The noise ratio of the estimation process on the $x$ position and $p$ momentum quadrature components separately can be identified by $\vartheta$ as

$$\vartheta = \frac{\left((\chi(\mathbf{C}))^T \chi(\mathbf{q}_j)\right)^2 \mathbb{E}\left[\chi(A_j)^2\right]}{|\chi(\mathbf{C})|^2 \sigma_\mathcal{N}^2}, \quad (44)$$

where function $\chi(\cdot)$ separately identifies the $x$ position or $p$ momentum quadrature components as $\mathrm{Re}(\cdot)$ and $\mathrm{Im}(\cdot)$, respectively. Note that

$$\left((\chi(\mathbf{C}))^T \chi(\mathbf{q}_j)\right)^2 \leq |\chi(\mathbf{C})|^2 |\chi(\mathbf{q}_j)|^2, \quad (45)$$

by some fundamental theory [23-25].

The $p_{err}(p'_x)$ of a noisy pilot-subcarrier CV $p'_x$ determines the noise of $\zeta(F(T_i(\mathcal{N}_i)))$. Since the $p_{err}(p'_x)$ quantifies the amount of noise $F(\Delta)$ on $p'_x$ and $\mathcal{S}_i(\mathcal{N}_i)$ (see (22), (24)), we derive this quantity by $\zeta(F(T_i(\mathcal{N}_i)))$ via an exact formula, as follows.

Let assume the case that $p_i = \mathrm{Re}(p_i) + \mathrm{Im}(p_i) = x_{p_i} + \mathrm{i}p_{p_i} \in \mathcal{C}$, $|p_i| > 0$ is a constant $p_i = p_x$ complex variable (CV quantum state with position $x_{p_i}$ and momentum quadrature $p_{p_i}$) for all $\mathcal{N}_i, i = 0, \ldots, l-1$ (which is reasonable in a practical CVQKD calibration phase), the error probability of the detection of the noisy $p'_x$ pilot-subcarrier CVs is evaluated as follows.

Let $\mathbf{p}_x \in \mathcal{C}^l$ identify an $l$-dimensional complex pilot-subcarrier CV vector,
$$\mathbf{p}_x = (p_{x,0}, \ldots, p_{x,l-1})^T, \tag{46}$$
where $p_{x,i}$ is the pilot-subcarrier CV of sub-channel $\mathcal{N}_i, i = 0, \ldots, l-1$, such that
$$p_{x,i} = p_x \in \mathcal{C}, i = 0, \ldots, l-1, \tag{47}$$
and $\mathbf{p}'_x$ the output vector as $\mathbf{p}'_x = (p'_{x,0}, \ldots, p'_{x,l-1})^T$.

Let $p_i = p_{x,i}$ for all $i$, then
$$\mathbf{p}'_x = \zeta(F(\mathbf{v}(\mathcal{N})))\mathbf{p}_x + F(\mathbf{v}(\Delta)), \tag{48}$$
which can be rewritten as
$$\mathbf{p}'_x = \zeta(F(\mathbf{v}(\mathcal{N})))p_x + F(\mathbf{v}(\Delta)), \tag{49}$$
where $\zeta(F(\mathbf{v}(\mathcal{N})))$ is expressed as
$$\zeta(F(\mathbf{v}(\mathcal{N}))) = (\zeta(F(T(\mathcal{N}_0))), \ldots, \zeta(F(T(\mathcal{N}_{l-1}))))^T, \tag{50}$$
where $\mathbf{v}(\mathcal{N})$ is the $l$-dimensional sub-channel vector.

Without loss of generality, the $\mathcal{S}(p'_x)$ sufficient statistics for the detection of $p'_x$ is then yielded [23] precisely as
$$\begin{aligned}\mathcal{S}(p'_x) &= \eta^\dagger \mathbf{p}'_x \\ &= \frac{\zeta(F(\mathbf{v}(\mathcal{N})))^\dagger}{|\zeta(F(\mathbf{v}(\mathcal{N})))|}\mathbf{p}'_x \\ &= |\zeta(F(\mathbf{v}(\mathcal{N})))|p_x + \frac{\zeta(F(\mathbf{v}(\mathcal{N})))^\dagger}{|\zeta(F(\mathbf{v}(\mathcal{N})))|}F(\mathbf{v}(\Delta)),\end{aligned} \tag{51}$$
where
$$\eta = \frac{\zeta(F(\mathbf{v}(\mathcal{N})))}{|\zeta(F(\mathbf{v}(\mathcal{N})))|}, \tag{52}$$
and
$$\eta^\dagger F(\mathbf{v}(\Delta)) \in \mathcal{CN}(0, 2\sigma_\mathcal{N}^2). \tag{53}$$

Specifically, assuming that each $\zeta(T(\mathcal{N}_i))$ of $\zeta(\mathbf{v}(\mathcal{N}))$ has a distribution of
$$\zeta(T(\mathcal{N}_i)) \in \mathcal{CN}(0, \sigma_{\zeta(T(\mathcal{N}_i))}^2), \tag{54}$$

the quantity of $\left|\zeta\left(F\left(\mathbf{v}\left(\mathcal{N}\right)\right)\right)\right|^2$ is $\chi^{2l}$ Chi-square distributed [23-25] with $2l$ degrees of freedom, which yields a density $f^{\chi^{2l}}(x)$ precisely as

$$f^{\chi^{2l}}(x) = \frac{1}{(l-1)!} x^{l-1} e^{-x}, \tag{55}$$

where $x \geq 0$. In particular, the density of (55) at $x \to 0$ is simplified into

$$f^{\chi^{2l}}(x) = \frac{1}{(l-1)!} x^{l-1}. \tag{56}$$

From (55), the $p_{err}(p'_x)$ error probability of the detection of $p'_x$ for the $l$ sub-channels is

$$\begin{aligned} p_{err}(p'_x) &= \int_0^\infty Q\left(\sqrt{2x\widehat{\mathrm{SNR}}}\right) f^{\chi^{2l}}(x) dx \\ &= \left(\frac{1-\sqrt{\frac{\widehat{\mathrm{SNR}}}{1+\widehat{\mathrm{SNR}}}}}{2}\right)^l \sum_{i=0}^{l-1} \binom{l-1+i}{i} \left(\frac{1+\sqrt{\frac{\widehat{\mathrm{SNR}}}{1+\widehat{\mathrm{SNR}}}}}{2}\right)^i, \end{aligned} \tag{57}$$

where $Q(\cdot)$ is the Gaussian tail function [23], and $\widehat{\mathrm{SNR}}$ is a scaled SNR quantity as

$$\widehat{\mathrm{SNR}} = \frac{0.5|p_x|^2}{2\sigma_\mathcal{N}^2} = \frac{1}{2}\mathrm{SNR}, \tag{58}$$

where SNR is the complex SNR.

The $p_{err}\left(p'_x \middle| \zeta\left(F\left(\mathbf{v}\left(\mathcal{N}\right)\right)\right)\right)$ conditional error probability is evaluated as [23]

$$p_{err}\left(p'_x \middle| \zeta\left(F\left(\mathbf{v}\left(\mathcal{N}\right)\right)\right)\right) = Q\sqrt{2\left|\zeta\left(F\left(\mathbf{v}\left(\mathcal{N}\right)\right)\right)\right|^2 \widehat{\mathrm{SNR}}}. \tag{59}$$

In particular, the result in (57) at $\widehat{\mathrm{SNR}} > 1$ is simplified into

$$p_{err}(p'_x) = \left(\frac{1}{4\widehat{\mathrm{SNR}}}\right)^l \binom{2l-1}{l}. \tag{60}$$

Introducing an error event

$$\mathrm{E}: \left|\zeta\left(F\left(\mathbf{v}\left(\mathcal{N}\right)\right)\right)\right|^2 < \frac{1}{\widehat{\mathrm{SNR}}}, \tag{61}$$

the corresponding $\Pr(\mathrm{E})$ is

$$\Pr(\mathrm{E}) \approx p_{err}(p'_x), \tag{62}$$

which is at $x \to 0$ without loss of generality is yielded as

$$\begin{aligned} \Pr(\mathrm{E}) &= \int_0^{1/\widehat{\mathrm{SNR}}} f^{\chi^{2l}}(x) dx \\ &= \int_0^{1/\widehat{\mathrm{SNR}}} \frac{1}{(l-1)!} x^{l-1} dx \\ &= \frac{1}{l!} \frac{1}{\widehat{\mathrm{SNR}}^l}. \end{aligned} \tag{63}$$

For the subcarrier spreading-based sub-channel estimation scheme with a theoretical minimum of $p_{err}$ in the presence of a Gaussian noise, see Lemma 1.

∎

## 3.2 Sub-channel Estimation with Subcarrier Spreading

**Lemma 1** (Subcarrier spreading-based sub-channel estimation for multicarrier CVQKD). *The subcarrier spreading results in a minimized sub-channel estimation error $p_{err} = Q\left(\sqrt{\Upsilon/2\sigma_{\mathcal{N}}^2}\right)$, where $\Upsilon = \left|\mathbf{p}_x\right|^2 \!/ g$, and $\mathbf{p}_x \in \mathcal{C}^g$ is $g$-dimensional pilot-subcarrier CV vector.*

*Proof.*
In Theorem 1, it has been already shown that the $p_{err}\left(p'_x\right)$ error probability quantifies the amount of noise on the $\mathcal{S}$ sufficient statistics. Here we prove that the error of the sub-channel estimation can be minimized via a technique, called subcarrier spreading. This type of transmission technique allows to evaluate $\zeta\left(F\left(T\left(\mathcal{N}_i\right)\right)\right)$ with minimal error.

Let $\mathbf{q}_x \in \mathcal{C}^g$ be a $g$-dimensional pilot-subcarrier CV vector

$$\mathbf{q}_x = \left(p_{x,0},\ldots,p_{x,g-1}\right)^T, \tag{64}$$

where

$$g + \left(l - 1\right) = n, \tag{65}$$

and $p_{x,i}$ is sent through sub-channel $\mathcal{N}_i$, $i = 0,\ldots,l-1$, $l < n$.

Then define $n$-dimensional vector $\mathbf{P}_x^i \in \mathcal{C}^n$, $i = 0,\ldots,l-1$ precisely as

$$\mathbf{P}_x^i = \left(\mathbf{a}_0, \mathbf{q}_x, \mathbf{b}_0\right), \tag{66}$$

where $\mathbf{a}_0$ is an $i$-dimensional vector

$$\mathbf{a}_0 = \left(\rho_0^0,\ldots,\rho_{i-1}^0\right)^T, \tag{67}$$

where $\rho_w^0 = \left|0\right\rangle\!\left\langle 0\right|, w = 0,\ldots,i-1$, and $\mathbf{b}_0$ is an $u = \left(l-1\right) - i$-dimensional vector

$$\mathbf{b}_0 = \left(\rho_0^0,\ldots,\rho_{u-1}^0\right)^T, \tag{68}$$

where $\rho_w^0 = \left|0\right\rangle\!\left\langle 0\right|, w = 0,\ldots,u-1$.

Then for any $i$, without loss of generality,

$$\left|\left(\mathbf{P}_x^i\right)^\dagger \left(\mathbf{P}_x^m\right)\right| < \left|\mathbf{q}_x\right|^2 = \sum_{i=0}^{g-1}\left|p_{x,i}\right|^2, \tag{69}$$

where $m \neq i$. Assuming full orthogonality [23], for any $m \neq i$,

$$\left(\mathbf{P}_x^i\right)^\dagger \left(\mathbf{P}_x^m\right) = 0. \tag{70}$$

Then let output of the $i$-th iteration is

$$\mathbf{P}_x^{i\prime} = F\left(T_i\left(\mathcal{N}_i\right)\right)\mathbf{P}_x^i + F\left(\mathbf{V}^i\left(\Delta\right)\right) \tag{71}$$

where $\mathcal{N}_i$ is the $i$-th sub-channel of the total $l$ "good" sub-channels, and $F\left(\mathbf{V}^i\left(\Delta\right)\right)$ is an $n$-dimensional vector.

The $\mathbf{P}_x^{i\prime}$ outputs of the $l$ iterations formulate the output vector $\mathbf{P}_{out}$ as

$$\begin{aligned}\mathbf{P}_{out} &= \sum_{l}\mathbf{P}_x^{i\prime} \\ &= \sum_{l} F\left(T_i\left(\mathcal{N}_i\right)\right)\mathbf{P}_x^i + F\left(\mathbf{V}^i\left(\Delta\right)\right).\end{aligned} \quad (72)$$

After some calculations, $p_{err}$ is yielded as

$$p_{err} = \mathbb{E}\left[Q\left(\sqrt{\frac{|\mathbf{q}_x|^2\left(\sum_{i=0}^{l-1}|F(T_i(\mathcal{N}_i))|^2\right)}{g2\sigma_{\mathcal{N}}^2}}\right)\right]. \quad (73)$$

Assuming that $p_{x,i} = p_x \in \mathcal{C}, i = 0,\dots,g-1$ holds in (64), the vector $\mathbf{q}_x$ can be rewritten as a $\mathbf{p}_x$ vector,

$$\mathbf{p}_x = \left(p_{x,0},\dots,p_{x,g-1}\right)^T, \quad (74)$$

and further assuming that each $F\left(T\left(\mathcal{N}_i\right)\right)$ has a distribution of $F\left(T\left(\mathcal{N}_i\right)\right) \in \mathcal{CN}\left(0, \sigma_{T(\mathcal{N}_i)}^2\right)$, (73) results in

$$p_{err}\left(p_x'\right) = \left(\frac{1-\sqrt{\frac{\widehat{\mathrm{SNR}}}{1+\widehat{\mathrm{SNR}}}}}{2}\right)^l \sum_{i=0}^{l-1}\binom{l-1+i}{i}\left(\frac{1+\sqrt{\frac{\widehat{\mathrm{SNR}}}{1+\widehat{\mathrm{SNR}}}}}{2}\right)^i, \quad (75)$$

which coincidences with (57), and

$$\widehat{\mathrm{SNR}} = \frac{0.5|p_x|^2}{2\sigma_{\mathcal{N}}^2} = \frac{1}{2}\mathrm{SNR}. \quad (76)$$

From the law of large numbers, for $l \to \infty$ follows that

$$\mathbb{E}\left[\sum_l |F\left(T\left(\mathcal{N}_i\right)\right)|^2\right] = 1 \quad (77)$$

with unit probability, which allows to rewrite (73) specifically as

$$\begin{aligned}p_{err} &= \mathbb{E}\left[Q\left(\sqrt{\frac{|\mathbf{p}_x|^2\left(\sum_{i=0}^{l-1}|F(T_i(\mathcal{N}_i))|^2\right)}{g2\sigma_{\mathcal{N}}^2}}\right)\right] \\ &= \mathbb{E}\left[Q\left(\sqrt{\frac{|p_x|^2\left(\sum_{i=0}^{l-1}|F(T_i(\mathcal{N}_i))|^2\right)}{2\sigma_{\mathcal{N}}^2}}\right)\right] \\ &= Q\left(\sqrt{\Upsilon\frac{1}{2\sigma_{\mathcal{N}}^2}}\right) \\ &= Q\left(\sqrt{2\widehat{\mathrm{SNR}}}\right) \\ &= Q\left(\sqrt{\mathrm{SNR}}\right),\end{aligned} \quad (78)$$

where

$$\Upsilon = \frac{|\mathbf{p}_x|^2}{g} = |p_x|^2. \quad (79)$$

In particular, projecting $\mathbf{P}_x^{i\prime}$ onto $\varsigma_x = \mathbf{P}_x^i/|\mathbf{P}_x^i|$ results $\mathcal{S}\left(\mathcal{N}_i\right)$ as [23]

$$\mathcal{S}(\mathcal{N}_i) = \varsigma_x^\dagger \mathbf{P}_x^{i\prime}$$
$$= \frac{\left(\mathbf{P}_x^i\right)^\dagger}{\left|\mathbf{P}_x^i\right|}\left(F(T_i(\mathcal{N}_i))\mathbf{P}_x^i + F(\mathbf{V}^i(\Delta))\right) \quad (80)$$
$$= F(T_i(\mathcal{N}_i))\left|\mathbf{P}_x^i\right| + F(\Delta).$$

The subcarrier spreading technique is illustrated in Fig. 2 for $n=5, l=3, g=3$. The subcarrier spreading technique scans through the $n$ sub-channels to estimate the transmittance coefficients of the $l$ "good" sub-channels $\mathcal{N}_0, \mathcal{N}_1$, and $\mathcal{N}_4$ (depicted by blue). The process iterates in $l$ steps. The scanned sub-channels of the $i$-th iteration step are depicted by the thick frame.

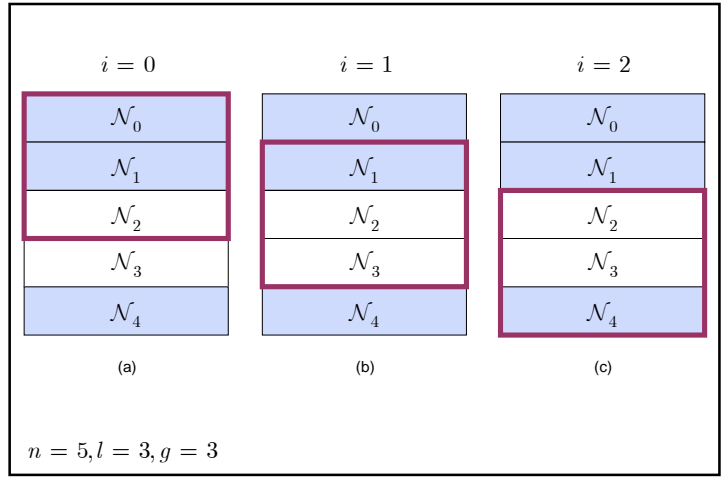

**Figure 2.** The subcarrier spreading technique at $n=5, l=3, g=3$. The $n$ sub-channels are scanned over in $l$ steps (a–c), $i=0,\ldots,l-1$, utilizing $g$ pilot-subcarrier CVs (sliding thick frame) in each iteration steps (The "good" sub-channels are depicted in blue.).

From (80), the $\zeta(F(T_i(\mathcal{N}_i)))$ linear estimation of $F(T_i(\mathcal{N}_i))$ is expressed as

$$\zeta(F(T_i(\mathcal{N}_i))) = \frac{\left|\mathbf{P}_x^i\right|}{\left|\mathbf{P}_x^i\right|^2 + 2\sigma_\mathcal{N}^2} \mathcal{S}(\mathcal{N}_i), \quad (81)$$

which yields a $\mathcal{E}$ mean square error without loss of generality as

$$\mathcal{E} = \mathbb{E}\left[\frac{1}{1+\left|\mathbf{P}_x^i\right|^2 \frac{1}{2\sigma_\mathcal{N}^2}}\right] \quad (82)$$

and

$$\widehat{\mathrm{SNR}} = \frac{0.5\left|\mathbf{P}_x^i\right|^2}{g 2\sigma_\mathcal{N}^2} = \frac{0.5|p_x|^2}{2\sigma_\mathcal{N}^2}. \quad (83)$$

Assuming that the sub-channel estimation phase is repeated for $k$-times, the resulting $\zeta^{(k)}(F(T_i(\mathcal{N}_i)))$ is

$$\zeta^{(k)}\big(F\big(T_i\big(\mathcal{N}_i\big)\big)\big) = \frac{\big|\mathbf{P}_x^i\big|}{k\big|\mathbf{P}_x^i\big|^2 + 2\sigma_{\mathcal{N}}^2} \sum_k \mathcal{S}\big(\mathcal{N}_i\big), \tag{84}$$

from which

$$\mathcal{E}^{(k)} = \mathbb{E}\left[\frac{1}{1 + k\big|\mathbf{P}_x^i\big|^2 \frac{1}{2\sigma_{\mathcal{N}}^2}}\right], \tag{85}$$

and

$$\widehat{\mathrm{SNR}}^{(k)} = \frac{0.5k\big|\mathbf{P}_x^i\big|^2}{g 2\sigma_{\mathcal{N}}^2} = \frac{0.5k\big|p_x\big|^2}{2\sigma_{\mathcal{N}}^2}. \tag{86}$$

In particular, the resulting error probability is

$$\begin{aligned}
p_{err}^{(k)} &= Q\!\left(\sqrt{2\widehat{\mathrm{SNR}}^{(k)}}\right) \\
&= Q\!\left(\sqrt{k\big|p_x\big|^2/2\sigma_{\mathcal{N}}^2}\right) \\
&= Q\!\left(\sqrt{k\mathrm{SNR}}\right).
\end{aligned} \tag{87}$$

Hence, the subcarrier spreading technique allows us to construct $\zeta\big(F\big(T\big(\mathcal{N}_i\big)\big)\big)$ of $F\big(T\big(\mathcal{N}_i\big)\big)$ with a theoretical error-minimum [23-25] in the presence of a Gaussian noise.

The $p_{err}$ and $p_{err}^{(k)}$ error probabilities of (57) and (87) for $k = 2$ at $F\big(T\big(\mathcal{N}_i\big)\big) \in \mathcal{CN}\big(0, \sigma_{T(\mathcal{N}_i)}^2\big)$, are compared in Fig. 3. By an averaging over $\mathcal{CN}\big(0, \sigma_{T(\mathcal{N}_i)}^2\big)$, the results can be extended for arbitrary distributed $F\big(T\big(\mathcal{N}_i\big)\big)$ sub-channel coefficients.

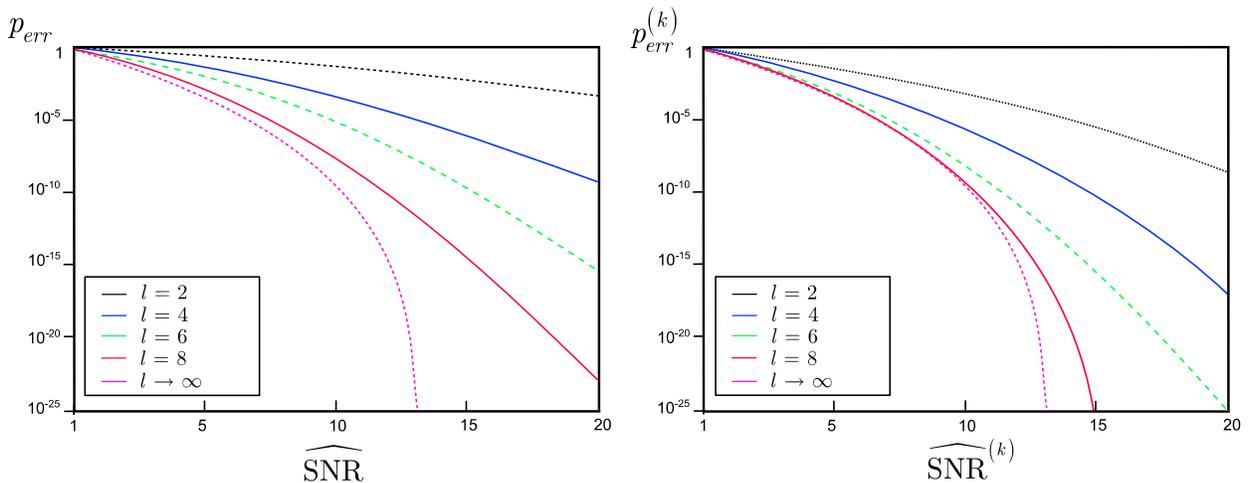

**Figure 3.** Comparison of the error probabilities of the sub-channel estimation ($p_{err}$) and the subcarrier spreading-based sub-channel estimation ($p_{err}^{(k)}$), at $k = 1,2$, $l = 2,4,6,8$ and $l \to \infty$, $F\big(T\big(\mathcal{N}_i\big)\big) \in \mathcal{CN}\big(0, \sigma_{T(\mathcal{N}_i)}^2\big)$.

∎

# 4 Adaptive Detection for Multicarrier CVQKD

First the results are proposed for single adaptive quadrature detection. The collective adaptive quadrature detection for multicarrier CVQKD is discussed in the second part.

## 4.1 Single Adaptive Quadrature Detection

**Proposition 1** (Single adaptive multicarrier detection). *For any $z \in \mathcal{CN}\left(0, 2\sigma_{\omega_0}^2\right)$, the $\mathcal{S}$ for a single detection is $\mathcal{S} = \nu^{\dagger} z'$, where $\nu = A/|A|$, $A = \frac{1}{l}\sum_{i=0}^{l-1} F\left(T_i\left(\mathcal{N}_i\right)\right)$, $Az \in \mathcal{CN}\left(0, 2\sigma_{\omega_0}^2\right)$, $z' = Az + F(\Delta) \in \mathcal{CN}\left(0, 2\left(\sigma_{\omega_0}^2 + \sigma_{\mathcal{N}}^2\right)\right)$, and $\nu^{\dagger} F(\Delta) \in \mathcal{CN}\left(0, 2\sigma_{\mathcal{N}}^2\right)$.*

*Proof.*
The first part of the proof discusses the $M_{\text{hom}}$ homodyne detection. In the second part, the results for the $M_{\text{het}}$ heterodyne measurement setting are derived.

Let $\chi(z')$, $z' \in \mathcal{CN}\left(0, 2\left(\sigma_{\omega_0}^2 + \sigma_{\mathcal{N}}^2\right)\right)$ identify the result of an $M_{\text{hom}}$ single homodyne measurement applied on the output Gaussian CV $\left|\phi_j'\right\rangle$ as

$$x', p' \in \mathbb{N}\left(0, \sigma_{\omega_0}^2 + \sigma_{\mathcal{N}}^2\right), \tag{88}$$

where $x', p'$ stands for the position and momentum quadrature of $z' = x' + \mathrm{i}p'$.

The term $\chi(z')$ can be rewritten as

$$\chi(z') = \left(\tfrac{1}{l}\sum_{i=0}^{l-1}\left|F\left(\chi\left(T_i\left(\mathcal{N}_i\right)\right)\right)\right|\right)\chi(z) \in \mathbb{N}\left(0, \sigma_{\omega_0}^2 + \sigma_{\mathcal{N}}^2\right), \tag{89}$$

and $\nu$ can be defined as

$$\chi(\nu) = \frac{\chi(A)}{|\chi(A)|} = \frac{\left(\tfrac{1}{l}\sum_{i=0}^{l-1} F\left(\chi\left(T_i(\mathcal{N}_i)\right)\right)\right)}{\tfrac{1}{l}\sum_{i=0}^{l-1}\left|F\left(\chi\left(T_i(\mathcal{N}_i)\right)\right)\right|}. \tag{90}$$

In particular, from $\chi(z')$ and $\chi(\nu)$, the $\mathcal{S} = \left(\chi(\nu)\right)^T \chi(z')$ sufficient statistics [23] in $\mathbb{R}$ is as follows:

$$\begin{aligned}
\mathcal{S} &= \left(\chi(\nu)\right)^T \chi(z') \\
&= \frac{(\chi(A))^T}{|\chi(A)|}\chi(z') \\
&= \left(\tfrac{1}{l}\sum_{i=0}^{l-1}\left|F\left(\chi\left(T_i\left(\mathcal{N}_i\right)\right)\right)\right|\right)\chi(z') + \frac{\left(\tfrac{1}{l}\sum_{i=0}^{l-1} F\left(\chi\left(T_i(\mathcal{N}_i)\right)\right)\right)^{\dagger}}{\tfrac{1}{l}\sum_{i=0}^{l-1}\left|F\left(\chi\left(T_i(\mathcal{N}_i)\right)\right)\right|}\chi\left(F(\Delta)\right),
\end{aligned} \tag{91}$$

where $\chi\left(F(\Delta)\right) \in \mathbb{N}\left(0, \sigma_{\mathcal{N}}^2\right)$.

For $M_{\text{het}}$, let the output be identified via operator $\Lambda(\cdot)$ as

$$\Lambda(z') \in \mathcal{CN}\left(0, 2c\left(\sigma_{\omega_0}^2 + \sigma_{\mathcal{N}}^2\right)\right), \tag{92}$$

where $c$ is a real variable, and without loss of generality let $F\left(\Lambda\left(T_i\left(\mathcal{N}_i\right)\right)\right) \in \mathcal{CN}\left(0, c2\sigma_\mathcal{N}^2\right)$, and

$$\mathcal{U} = \frac{\Lambda(A)}{|\Lambda(A)|} = \frac{\left(\frac{1}{l}\sum_{i=0}^{l-1} F(\Lambda(T_i(\mathcal{N}_i)))\right)}{\frac{1}{l}\sum_{i=0}^{l-1}|F(\Lambda(T_i(\mathcal{N}_i)))|}; \tag{93}$$

thus,

$$\begin{aligned}\mathcal{S} &= \mathcal{U}^\dagger \Lambda(z') \\ &= \frac{\Lambda(A)^\dagger}{|\Lambda(A)|} \Lambda(z') \\ &= \left(\frac{1}{l}\sum_{i=0}^{l-1}\left|F\left(\Lambda\left(T_i\left(\mathcal{N}_i\right)\right)\right)\right|\right)\Lambda(z') + \frac{\left(\frac{1}{l}\sum_{i=0}^{l-1} F(\Lambda(T_i(\mathcal{N}_i)))\right)^\dagger}{\frac{1}{l}\sum_{i=0}^{l-1}|F(\Lambda(T_i(\mathcal{N}_i)))|}\Lambda\left(F(\Delta)\right), \end{aligned} \tag{94}$$

where $\Lambda\left(F(\Delta)\right) \in \mathcal{CN}\left(0, c2\sigma_\mathcal{N}^2\right)$.

The result in (94), at $c = 1$, can be rewritten as

$$\begin{aligned}\mathcal{S} &= \nu^\dagger z' \\ &= \frac{A^\dagger}{|A|} z' \\ &= \left(\frac{1}{l}\sum_{i=0}^{l-1}\left|F\left(T_i\left(\mathcal{N}_i\right)\right)\right|\right)z + \frac{\left(\frac{1}{l}\sum_{i=0}^{l-1} F(T_i(\mathcal{N}_i))\right)^\dagger}{\frac{1}{l}\sum_{i=0}^{l-1}|F(T_i(\mathcal{N}_i))|} F(\Delta), \end{aligned} \tag{95}$$

which provides $\mathcal{S} = \mathcal{U}^\dagger z'$ in the $\mathcal{C}$ complex space.

Note that if $z' = x' + \mathrm{i}p'$, the condition

$$x' \in \mathbb{N}\left(0, \sigma_{\omega_0}^2 + \sigma_\mathcal{N}^2\right) = p' \in \mathbb{N}\left(0, \sigma_{\omega_0}^2 + \sigma_\mathcal{N}^2\right) \tag{96}$$

is satisfied, then $\mathcal{S}$ can also be extracted in $\mathbb{R}$, using $\chi\left(F(\Delta)\right) \in \mathbb{N}\left(0, \sigma_\mathcal{N}^2\right) \in \mathbb{R}$, as shown in (91).

∎

The results are proposed for collective adaptive quadrature detection in Theorem 2.

## 4.2 Collective Adaptive Quadrature Detection

**Theorem 2** (Sufficient statistics for collective adaptive detection). *For a d-dimensional input* $\mathbf{z} = diag\left(z_0, \ldots, z_{d-1}\right) \in \mathcal{CN}\left(0, \mathbf{K_z}\right)$, *the sufficient statistics is* $\mathcal{S}^d = \left(\nu^d\right)^\dagger \mathbf{z}'$, *where* $\nu^d = \mathbf{A}^\dagger \mathbf{M}^N$, $\mathbf{A} = \left[\frac{1}{l}\sum_{i=0}^{l-1} F\left(T_{0,i}\left(\mathcal{N}_{0,i}\right)\right)^\dagger, \ldots, \frac{1}{l}\sum_{i=0}^{l-1} F\left(T_{d-1,i}\left(\mathcal{N}_{d-1,i}\right)\right)^\dagger\right]^T$, $\mathbf{M}^N$ *is the codeword difference matrix of the $N$ matrices* $\mathbf{z}_0, \ldots, \mathbf{z}_{N-1}$, $\mathbf{z}'^T = \mathbf{A}^\dagger \mathbf{z} + \left(F^d(\Delta)\right)^T \in \mathcal{CN}\left(0, \mathbf{K_{z'}}\right)$, *where* $F^d(\Delta) = \left(F(\Delta_0), \ldots, F(\Delta_{d-1})\right)^T \in \mathcal{CN}\left(0, \mathbf{K}_{F^d(\Delta)}\right)$, $|\mathbf{A}| = \sum_{j=0}^{d-1}\frac{1}{l}\sum_{i=0}^{l-1}\left|F\left(T_{j,i}\left(\mathcal{N}_{j,i}\right)\right)\right|$, $|\mathbf{A}|\mathbf{z} \in \mathcal{CN}\left(0, \mathbf{K_z}\right)$.

*Proof.*

Let the $d$-dimensional output $\mathbf{z}'$ be given as
$$\mathbf{z}'^T = \mathbf{A}^\dagger \mathbf{z} + \left(F^d(\Delta)\right)^T = \left(z'_0, \ldots, z'_{d-1}\right), \tag{97}$$

where $\mathbf{z} = diag(z_0, \ldots, z_{d-1}) \in \mathcal{CN}(0, \mathbf{K}_\mathbf{z})$ is a $d$-dimensional diagonal input matrix with $z_j \in \mathcal{CN}\left(0, 2\sigma^2_{\omega_0}\right)$, and $\mathbf{A}$ is a $d$-dimensional vector with the $\frac{1}{l}\sum_{i=0}^{l-1} F\left(T_{j,i}\left(\mathcal{N}_{j,i}\right)\right)$ averaged Fourier-transformed sub-channel coefficients for $z_j$, $j = 0, \ldots, d-1$, at $l$ sub-channels $\mathcal{N}_{j,i}, i = 0, \ldots, l-1$, as

$$\mathbf{A} = \left(A_0^\dagger, \ldots, A_{d-1}^\dagger\right)^T = \left[\frac{1}{l}\sum_{i=0}^{l-1} F\left(T_{0,i}\left(\mathcal{N}_{0,i}\right)\right)^\dagger, \ldots, \frac{1}{l}\sum_{i=0}^{l-1} F\left(T_{d-1,i}\left(\mathcal{N}_{d-1,i}\right)\right)^\dagger\right]^T, \tag{98}$$

and
$$F^d(\Delta) = \left(F(\Delta_0), \ldots, F(\Delta_{d-1})\right)^T \in \mathcal{CN}\left(0, \mathbf{K}_{F^d(\Delta)}\right), \tag{99}$$

$$z'_j = \left(\frac{1}{l}\sum_{i=0}^{l-1} F\left(T_{j,i}\left(\mathcal{N}_{j,i}\right)\right)\right) z_j + F(\Delta) \in \mathcal{CN}\left(0, 2\left(\sigma^2_{\omega_0} + \sigma^2_\mathcal{N}\right)\right). \tag{100}$$

From $\mathbf{A}$, $|\mathbf{A}|$ is expressed as

$$|\mathbf{A}| = \sum_{j=0}^{d-1} |A_j| = \sum_{j=0}^{d-1} \frac{1}{l}\sum_{i=0}^{l-1} \left|F\left(T_{j,i}\left(\mathcal{N}_{j,i}\right)\right)\right|; \tag{101}$$

thus,

$$|\mathbf{A}|\mathbf{z} = \sum_{j=0}^{d-1} \frac{1}{l}\sum_{i=0}^{l-1} \left|F\left(T_{j,i}\left(\mathcal{N}_{j,i}\right)\right)\right|\mathbf{z} \in \mathcal{CN}(0, \mathbf{K}_\mathbf{z}), \tag{102}$$

whereas $\mathbf{z}'^T$ is evaluated as

$$\begin{aligned}\mathbf{z}'^T &= \left(\frac{1}{l}\sum_{i=0}^{l-1} F\left(T_{0,i}\left(\mathcal{N}_{0,i}\right)\right), \ldots, \frac{1}{l}\sum_{i=0}^{l-1} F\left(T_{d-1,i}\left(\mathcal{N}_{d-1,i}\right)\right)\right) diag(z_0, \ldots, z_{d-1}) \\ &\quad + \left(F(\Delta_0), \ldots, F(\Delta_{d-1})\right) \\ &= \mathbf{A}^\dagger \mathbf{z} + \left(F^d(\Delta)\right)^T \in \mathcal{CN}(0, \mathbf{K}_{\mathbf{z}'}).\end{aligned} \tag{103}$$

Note that in a CVQKD setting, private classical information (i.e., $d$-dimensional random private classical $\mathbf{p}$ codewords) is shared between the legal parties. Thus, in particular, using $\mathbf{z}_A$ and $\mathbf{z}_B$, the $d$-dimensional random private codewords $\mathbf{p}_A$ and $\mathbf{p}_B$ [5-7] are defined with the relation
$$\mathcal{C}_{\mathbf{p}_A} \subseteq \mathcal{C}_{\mathbf{z}_A} \tag{104}$$

and
$$\mathcal{C}_{\mathbf{p}_B} \subseteq \mathcal{C}_{\mathbf{z}_B}, \tag{105}$$

where $\{\mathcal{C}_{\mathbf{z}_A}, \mathcal{C}_{\mathbf{z}_B}\}$ and $\{\mathcal{C}_{\mathbf{p}_A}, \mathcal{C}_{\mathbf{p}_B}\}$ are the corresponding phase-space constellations of $\mathbf{z}_A, \mathbf{z}_B$, and $\mathbf{p}_A, \mathbf{p}_B$, respectively. Because the results for $\mathbf{z}$ trivially follow for any $\mathbf{p}$, the proof demonstrates the decoding of $\mathbf{z}$.

Using (97), $\mathbf{z}$ at $N = 2$ is
$$\mathbf{A}^\dagger \mathbf{z} \in \left\{\mathbf{A}^\dagger \mathbf{z}_A, \mathbf{A}^\dagger \mathbf{z}_B\right\}, \tag{106}$$

where $\mathbf{z}_A = diag(z_{A,0},...,z_{A,d-1}) \in \mathcal{CN}(0, \mathbf{K}_{\mathbf{z}_A})$ and $\mathbf{z}_B = diag(z_{B,0},...,z_{B,d-1}) \in \mathcal{CN}(0, \mathbf{K}_{\mathbf{z}_B})$.

Without loss of generality, let the number of $d$-dimensional input codewords, $N = 2$, be denoted by $\mathbf{z}_A$ and $\mathbf{z}_B$. The results can be extended for arbitrary $N$.

Projecting $\mathbf{z}'$ onto the $d$-dimensional unit vector $\nu^d$, such that

$$\nu^d = \mathbf{A}^\dagger \mathbf{M}^2, \tag{107}$$

where $\mathbf{M}^2$ is the codeword difference matrix at $N = 2$, expressed as

$$\mathbf{M}^2 = \mathbf{z}_A - \mathbf{z}_B. \tag{108}$$

From (107) and (108), the $\mathcal{S}^d$ sufficient statistic is precisely as follows

$$\begin{aligned}\mathcal{S}^d &= \left(\nu^d\right)^\dagger \mathbf{z}' \\ &= s\left|\mathbf{A}^\dagger \mathbf{M}^2\right| + F(\Delta),\end{aligned} \tag{109}$$

where

$$s \in \{-0.5, 0.5\} \in \mathbb{R} \tag{110}$$

and $F(\Delta) \in \mathcal{CN}(0, 2\sigma_\mathcal{N}^2)$ are scalars.

To verify (109), first we rewrite (106) as

$$\mathbf{A}^\dagger \mathbf{z} = s\left(\mathbf{A}^\dagger \mathbf{z}_A - \mathbf{A}^\dagger \mathbf{z}_B\right) + \tfrac{1}{2}\left(\mathbf{A}^\dagger \mathbf{z}_A + \mathbf{A}^\dagger \mathbf{z}_B\right), \tag{111}$$

which lies in a $\gamma$ subspace of one $\mathbb{R}$ real dimension. From (111), let $\mathbf{Z}'$ be defined as

$$\begin{aligned}\mathbf{Z}'^T &= \mathbf{z}'^T - \tfrac{1}{2}\left(\mathbf{A}^\dagger \mathbf{z}_A + \mathbf{A}^\dagger \mathbf{z}_B\right) \\ &= \mathbf{A}^\dagger \mathbf{z} + \left(F^d(\Delta)\right)^T - \tfrac{1}{2}\left(\mathbf{A}^\dagger \mathbf{z}_A + \mathbf{A}^\dagger \mathbf{z}_B\right) \\ &= s\left(\mathbf{A}^\dagger \mathbf{z}_A - \mathbf{A}^\dagger \mathbf{z}_B\right) + \left(F^d(\Delta)\right)^T,\end{aligned} \tag{112}$$

where $F^d(\Delta) \in \mathcal{CN}(0, 2\sigma_\mathcal{N}^2 I)$.

Specifically, using the results obtained in (111) and (112), $\nu^d$ can be expressed as

$$\begin{aligned}\nu^d &= \frac{\mathbf{A}^\dagger \mathbf{z}_A - \mathbf{A}^\dagger \mathbf{z}_B}{\left|\left(\mathbf{A}^\dagger \mathbf{z}_A - \mathbf{A}^\dagger \mathbf{z}_B\right)\right|} \\ &= \frac{\mathbf{A}^\dagger(\mathbf{z}_A - \mathbf{z}_B)}{\left|\mathbf{A}^\dagger(\mathbf{z}_A - \mathbf{z}_B)\right|} = \frac{\mathbf{A}^\dagger \mathbf{M}^2}{\left|\mathbf{A}^\dagger \mathbf{M}^2\right|}.\end{aligned} \tag{113}$$

Thus, projecting $\mathbf{z}'$ along $\nu^d$ via $\mathcal{P}(\mathbf{z}')$ yields a complex scalar $\Gamma \in \mathcal{C}$ as follows:

$$\begin{aligned}\mathcal{P}(\mathbf{z}') &= \left(\nu^d\right)^\dagger \mathbf{Z}' \\ &= \left(\nu^d\right)^\dagger \left[\mathbf{z}' - \tfrac{1}{2}\left(\mathbf{A}^\dagger \mathbf{z}_A + \mathbf{A}^\dagger \mathbf{z}_B\right)^T\right] \\ &= s\left|\left(\mathbf{A}^\dagger \mathbf{z}_A - \mathbf{A}^\dagger \mathbf{z}_B\right)\right| + F(\Delta) \\ &= s\left|\mathbf{A}^\dagger(\mathbf{z}_A - \mathbf{z}_B)\right| + F(\Delta) \\ &= s\left|\mathbf{A}^\dagger \mathbf{M}^2\right| + F(\Delta) \\ &= \Gamma,\end{aligned} \tag{114}$$

where $F(\Delta) \in \mathcal{CN}(0, 2\sigma_\mathcal{N}^2)$ is a complex scalar.

Thus, $\text{Re}(\Gamma) \in \mathbb{R}$ and $\text{Im}(\Gamma) \in \mathbb{R}$ are real variables with noise

$$\chi\left(F\left(\Delta\right)\right) \in \mathbb{N}\left(0, \sigma_{\mathcal{N}}^2\right), \tag{115}$$

which identifies the $\text{Re}\left(F\left(\Delta\right)\right) \in \mathbb{N}\left(0, \sigma_{\mathcal{N}}^2\right)$ real or $\text{Im}\left(F\left(\Delta\right)\right) \in \mathbb{N}\left(0, \sigma_{\mathcal{N}}^2\right)$ imaginary component of $F\left(\Delta\right) \in \mathcal{CN}\left(0, 2\sigma_{\mathcal{N}}^2\right)$, respectively.

In particular, the fact that the result $\Gamma$ of the $\mathcal{P}$ projection contains all information for the decoding is verified as follows. Let $\mathbf{G}_{uv}$ be an orthogonal matrix [23], where $uv$ indexes the row and column of $\mathbf{G}$. Let $\mathbf{G}_{1v} \equiv \mathbf{A}$, and let the other unit norm rows to be orthogonal to $\mathbf{A}$ and to each other. Then, (114) yields a vector $\mathbf{L}$ precisely as

$$\mathbf{L} = \mathbf{G}\left(\mathbf{z}'^T - \tfrac{1}{2}\left(\mathbf{A}^\dagger \mathbf{z}_A + \mathbf{A}^\dagger \mathbf{z}_B\right)\right)^T = \begin{pmatrix} s\left|\left(\mathbf{A}^\dagger \mathbf{z}_A - \mathbf{A}^\dagger \mathbf{z}_B\right)\right| \\ 0 \\ \vdots \\ 0 \end{pmatrix} + \mathbf{G}F^d\left(\Delta\right), \tag{116}$$

where $\mathbf{G}F^d\left(\Delta\right) \in \mathcal{CN}\left(0, 2\sigma_{\mathcal{N}}^2 I\right)$, and $I$ is the $d \times d$ identity. From $\mathbf{L}$, only $s\left|\left(\mathbf{A}^\dagger \mathbf{z}_A - \mathbf{A}^\dagger \mathbf{z}_B\right)\right|$ (i.e., the first component of $\mathbf{L}$) is not independent from $s$ and $F^d\left(\Delta\right)$. Thus, all information is conveyed in the first component of $\mathbf{L}$, which exactly coincides with (114).

Exploiting some fundamentals of the maximum likelihood theory [23-25], the decision rule in $\mathcal{C}^d$ with respect to the $\mathbb{R}^d$ $d$-dimensional subspace with $\chi\left(F^d\left(\Delta\right)\right) \in \mathcal{CN}\left(0, \sigma_{\mathcal{N}}^2 I\right)$ is as

$$\mathbf{A}^\dagger \mathbf{z} = \mathbf{A}^\dagger \mathbf{z}_A : \frac{1}{\left(\pi 2\sigma_{\mathcal{N}}^2\right)^{d/2}} e^{\left(-\frac{\left|\mathbf{z}'^T - \mathbf{A}^\dagger \mathbf{z}_A\right|^2}{2\sigma_{\mathcal{N}}^2}\right)} \geq \frac{1}{\left(\pi 2\sigma_{\mathcal{N}}^2\right)^{d/2}} e^{\left(-\frac{\left|\mathbf{z}'^T - \mathbf{A}^\dagger \mathbf{z}_B\right|^2}{2\sigma_{\mathcal{N}}^2}\right)}, \tag{117}$$

and

$$\mathbf{A}^\dagger \mathbf{z} = \mathbf{A}^\dagger \mathbf{z}_B : \frac{1}{\left(\pi 2\sigma_{\mathcal{N}}^2\right)^{d/2}} e^{\left(-\frac{\left|\mathbf{z}'^T - \mathbf{A}^\dagger \mathbf{z}_B\right|^2}{2\sigma_{\mathcal{N}}^2}\right)} \geq \frac{1}{\left(\pi 2\sigma_{\mathcal{N}}^2\right)^{d/2}} e^{\left(-\frac{\left|\mathbf{z}'^T - \mathbf{A}^\dagger \mathbf{z}_A\right|^2}{2\sigma_{\mathcal{N}}^2}\right)}, \tag{118}$$

which conditions can be rewritten as

$$\mathbf{A}^\dagger \mathbf{z} = \mathbf{A}^\dagger \mathbf{z}_A : \left|\mathbf{z}'^T - \mathbf{A}^\dagger \mathbf{z}_A\right| < \left|\mathbf{z}'^T - \mathbf{A}^\dagger \mathbf{z}_B\right|, \tag{119}$$

and

$$\mathbf{A}^\dagger \mathbf{z} = \mathbf{A}^\dagger \mathbf{z}_B : \left|\mathbf{z}'^T - \mathbf{A}^\dagger \mathbf{z}_B\right| < \left|\mathbf{z}'^T - \mathbf{A}^\dagger \mathbf{z}_A\right|. \tag{120}$$

Precisely, using (114), the results of (117)–(118) with $\chi\left(F\left(\Delta\right)\right) \in \mathbb{N}\left(0, \sigma_{\mathcal{N}}^2\right)$ can be rewritten with respect to the subspace $\mathbb{R}$ of $\mathcal{C}$ as

$$Az = Az_A : \frac{1}{\sqrt{\pi 2\sigma_{\mathcal{N}}^2}} e^{\left(-\frac{\left(z' - \mu_{Az_A}\right)^2}{2\sigma_{\mathcal{N}}^2}\right)} \geq \frac{1}{\sqrt{\pi 2\sigma_{\mathcal{N}}^2}} e^{\left(-\frac{\left(z' - \mu_{Az_B}\right)^2}{2\sigma_{\mathcal{N}}^2}\right)}, \tag{121}$$

and

$$Az = Az_B : \frac{1}{\sqrt{\pi 2\sigma_\mathcal{N}^2}} e^{\left(-\frac{\left(z'-\mu_{Az_B}\right)^2}{2\sigma_\mathcal{N}^2}\right)} \geq \frac{1}{\sqrt{\pi 2\sigma_\mathcal{N}^2}} e^{\left(-\frac{\left(z'-\mu_{Az_A}\right)^2}{2\sigma_\mathcal{N}^2}\right)}, \tag{122}$$

whereas (119)–(120) are reevaluated as

$$Az = Az_A : \left|z' - \mu_{Az_A}\right| < \left|z' - \mu_{Az_B}\right|, \tag{123}$$

and

$$Az = Az_B : \left|z' - \mu_{Az_B}\right| < \left|z' - \mu_{Az_A}\right|. \tag{124}$$

The decision region is decomposed into hyperplanes $\mathcal{H}_{Az_A}$ and $\mathcal{H}_{Az_B}$ along $\nu^d$ via the decision rule $\frac{\left|\mathbf{A}^\dagger \mathbf{z}_A - \mathbf{A}^\dagger \mathbf{z}_B\right|}{2} = \frac{\mu_{Az_A} + \mu_{Az_B}}{2}$, where $\mu_{Az_A}, \mu_{Az_B} \in \mathcal{C}$ are the expected means of $\Gamma$, whereas $Az_A, Az_B \in \mathcal{C}$ and $z'_A, z'_B \in \mathcal{C}$ are complex scalar quantities.

The $\mathcal{P}(\mathbf{z}')$ projection of $\mathbf{z}'$ along $\nu^d$ is depicted in Fig. 4. If $\Gamma \in Az_A$, then $\mathbf{A}^\dagger \mathbf{z} = \mathbf{A}^\dagger \mathbf{z}_A$; if $\Gamma \in Az_B$, then $\mathbf{A}^\dagger \mathbf{z} = \mathbf{A}^\dagger \mathbf{z}_B$, where $\mathbf{A}^\dagger$ is obtained via the sub-channel estimation phase of Theorem 1. Finally, the results of Proposition 1 yield the decision on $\mathbf{z} = \mathbf{z}_A$ or $\mathbf{z} = \mathbf{z}_B$, respectively.

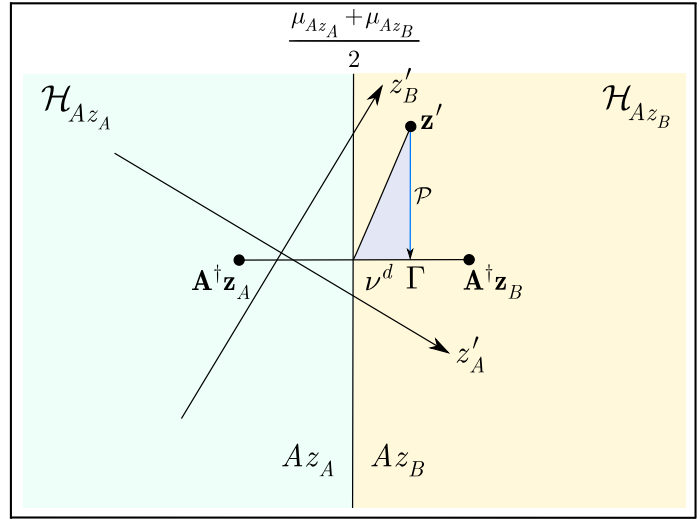

**Figure 4**. The $\mathcal{P}(\mathbf{z}')$ projection of $d$-dimensional $\mathbf{z}' \in \mathcal{C}^d$ along $\nu^d$ yields the $\mathcal{S}^d$ sufficient statistics as a complex scalar quantity $\Gamma \in \mathcal{C}$. $N = 2$.

The $p_{err}$ error probability in the $\mathcal{C}^d$ $d$-dimensional complex vector space is

$$\begin{aligned} p_{err} &= \Pr\left(\left|\left(F^d(\Delta)\right)\right|^2 > \left|\left(F^d(\Delta)\right)^T + \mathbf{A}^\dagger\mathbf{z}_A - \mathbf{A}^\dagger\mathbf{z}_B\right|^2\right) \\ &= \Pr\left(\left(\mathbf{A}^\dagger\mathbf{z}_A - \mathbf{A}^\dagger\mathbf{z}_B\right)F^d(\Delta) < -\frac{\left|\left(\mathbf{A}^\dagger\mathbf{z}_A-\mathbf{A}^\dagger\mathbf{z}_B\right)\right|^2}{2}\right) \\ &= Q\left(\frac{\left|\left(\mathbf{A}^\dagger\mathbf{z}_A-\mathbf{A}^\dagger\mathbf{z}_B\right)\right|}{2\sqrt{2}\sigma_\mathcal{N}}\right), \end{aligned} \tag{125}$$

where
$$\left(\mathbf{A}^\dagger \mathbf{z}_A - \mathbf{A}^\dagger \mathbf{z}_B\right) F^d\left(\Delta\right) \in \mathcal{CN}\left(0, \left|\left(\mathbf{A}^\dagger \mathbf{z}_A - \mathbf{A}^\dagger \mathbf{z}_B\right)\right|^2 2\sigma_\mathcal{N}^2\right). \tag{126}$$

In the $\mathbb{R}^d$ $d$-dimensional subspace of $\mathcal{C}^d$, it is evaluated with $\chi\left(F^d\left(\Delta\right)\right) \in \mathbb{N}\left(0, \sigma_\mathcal{N}^2 I\right)$ as

$$\begin{aligned}
p_{err} &= \Pr\left[\left|\chi\left(F^d\left(\Delta\right)\right)\right|^2 > \left|\chi\left(F^d\left(\Delta\right)\right)^T + \chi\left(\mathbf{A}^\dagger \mathbf{z}_A - \mathbf{A}^\dagger \mathbf{z}_B\right)\right|^2\right] \\
&= \Pr\left(\chi\left(\mathbf{A}^\dagger \mathbf{z}_A - \mathbf{A}^\dagger \mathbf{z}_B\right)\chi\left(F^d\left(\Delta\right)\right) < -\frac{\left|\chi\left(\mathbf{A}^\dagger \mathbf{z}_A - \mathbf{A}^\dagger \mathbf{z}_B\right)\right|^2}{2}\right) \\
&= Q\left(\frac{\left|\mathbf{A}^\dagger \mathbf{z}_A - \mathbf{A}^\dagger \mathbf{z}_B\right|}{2\sigma_\mathcal{N}}\right),
\end{aligned} \tag{127}$$

where
$$\chi\left(\mathbf{A}^\dagger \mathbf{z}_A - \mathbf{A}^\dagger \mathbf{z}_B\right)\chi\left(F^d\left(\Delta\right)\right) \in \mathbb{N}\left(0, \left|\chi\left(\mathbf{A}^\dagger \mathbf{z}_A - \mathbf{A}^\dagger \mathbf{z}_B\right)\right|^2 \sigma_\mathcal{N}^2\right). \tag{128}$$

Exploiting $\Gamma$, the $p_{err}$ decoding error probability in $\mathcal{C}$ is

$$\begin{aligned}
p_{err} &= \Pr\left(z' < \tfrac{1}{2}\left(\mu_{Az_A} + \mu_{Az_B}\right) \Big| z = Az_A\right) \\
&= \Pr\left(F\left(\Delta\right) > \tfrac{1}{2}\left(\left|\mu_{Az_A} - \mu_{Az_B}\right|\right)\right) \\
&= Q\left(\frac{\left|\mu_{Az_A} - \mu_{Az_B}\right|}{2\sqrt{2}\sigma_\mathcal{N}}\right),
\end{aligned} \tag{129}$$

and in the subspace of $\mathbb{R}$, it can be evaluated precisely as

$$\begin{aligned}
p_{err} &= \Pr\left(\chi\left(z'\right) < \tfrac{1}{2}\chi\left(\mu_{Az_A} + \mu_{Az_B}\right) \Big| \chi\left(z\right) = \chi\left(Az_A\right)\right) \\
&= \Pr\left(\chi\left(F\left(\Delta\right)\right) > \tfrac{1}{2}\left(\left|\chi\left(\mu_{Az_A} - \mu_{Az_B}\right)\right|\right)\right) \\
&= Q\left(\frac{\left|\mu_{Az_A} - \mu_{Az_B}\right|}{2\sigma_\mathcal{N}}\right).
\end{aligned} \tag{130}$$

Focusing on subspace $\mathbb{R}^d$ with $N$ codewords, at a given $\mathbf{A}$, let E be an error event. Then the conditional error probability is

$$\Pr\left(\mathrm{E}\big|\mathbf{A}\right) = Q\left(\frac{\left|\mathbf{A}^\dagger \mathbf{M}^N\right|}{2\sigma_\mathcal{N}}\right). \tag{131}$$

After some calculations, the corresponding error probability is yielded as

$$\Pr\left(\mathrm{E}\right) = \mathbb{E}\left[Q\left(\sqrt{\frac{\mathrm{SNR}\mathbf{A}^\dagger \mathbf{M}^N \left(\mathbf{M}^N\right)^\dagger \mathbf{A}}{2}}\right)\right]. \tag{132}$$

The matrix $\mathbf{M}^N$ is Hermitian, that is, $\left(\mathbf{M}^N\right)^\dagger = \mathbf{M}^N$, which can be diagonalized by a $U$ unitary operation, such that $\mathbf{M}^N \left(\mathbf{M}^N\right)^\dagger = U\tau U^\dagger$, where

$$\tau = diag\left(\lambda_0^2, \ldots, \lambda_{d-1}^2\right), \tag{133}$$

where $\lambda_i^2$ are the singular values of $\mathbf{M}^N$ [23-25].

In particular, the result in (133) can be further exploited to derive $\Pr\left(\mathrm{E}\right)$ as

$$\Pr(\mathrm{E}) = \mathbb{E}\left[Q\left(\sqrt{\frac{\mathrm{SNR}\left|U^{\dagger}\mathbf{A}\right|^{2}\tau}{2}}\right)\right]$$
$$= \mathbb{E}\left[Q\left(\sqrt{\frac{\mathrm{SNR}\sum_{j=0}^{d-1}\left|U_{j}^{\dagger}A_{j}\right|^{2}\lambda_{j}^{2}}{2}}\right)\right], \quad (134)$$

where an averaging over a $U_{j}^{\dagger}A_{j} \in \mathcal{CN}\left(0,\sigma_{A_{j}}^{2}\right)$ distribution leads to an upper bound on $\Pr(\mathrm{E})$ as

$$\Pr(\mathrm{E}) \leq \mathbb{E}\left[\prod_{j=0}^{d-1}\frac{1}{1+\mathrm{SNR}\frac{1}{4}\lambda_{j}^{2}}\right]. \quad (135)$$

Specifically, assuming the case of $N \geq d$, and $\forall \lambda_i^2 > 0$, after some calculations the success probability is yielded as precisely

$$\Pr \geq 1 - \mathbb{E}\left[\frac{4^d}{\mathrm{SNR}^d \prod_{j=0}^{d-1}\lambda_j^2}\right]$$
$$= 1 - \mathbb{E}\left[\frac{4^d}{\mathrm{SNR}^d \det\left(\mathbf{M}^N\left(\mathbf{M}^N\right)^{\dagger}\right)}\right]. \quad (136)$$

Putting the pieces together, the $\mathcal{S}^d$ sufficient statistics for the decoding of $d$-dimensional outputs in a multicarrier CVQKD scenario at $N = 2$ is

$$\mathcal{S}^d = \left(\nu^d\right)^{\dagger} \mathbf{z}'$$
$$= s\left|\mathbf{A}^{\dagger}\mathbf{M}^2\right| + F(\Delta), \quad (137)$$

which result can be extended for arbitrary $N$ as

$$\mathcal{S}^d = \left(\nu^d\right)^{\dagger} \mathbf{z}'$$
$$= \left(\frac{\mathbf{A}^{\dagger}\mathbf{M}^N}{\left|\mathbf{A}^{\dagger}\mathbf{M}^N\right|}\right)^{\dagger} \mathbf{z}', \quad (138)$$

where $\mathbf{M}^N$ is the codeword difference matrix of the $N$ codewords $\mathbf{z} \in \left\{\mathbf{z}_0,\ldots,\mathbf{z}_{N-1}\right\}$.

∎

## 5 Multiuser Adaptive Quadrature Detection

**Theorem 3** (Adaptive quadrature detection for multiuser multicarrier CVQKD). *Assuming a $\left(K_{in}, K_{out}\right)$ multiuser multicarrier CVQKD setting, the $p_{err}$ error probability of the adaptive detection of user $U_k$ is $p_{err} \leq \mathbb{E}\left[\prod_{j=0}^{r_k-1}\frac{1}{1+\mathrm{SNR}_{U_k}\frac{1}{4}\lambda_j^2}\right]$, where $\lambda_j^2$ is the j-th singular value of the $\mathbf{M}_{U_k}^N$ codeword difference matrix of the $N$, $r_k$-dimensional $\tilde{\mathbf{z}}_k \in \mathcal{CN}\left(0,\mathbf{K}_{\tilde{\mathbf{z}}_k}\right)$ codewords of $U_k$, $r_k \leq d$, $\sum_{K_{out}} r_k = d$, and $\mathrm{SNR}_{U_k}$ is the SNR of $U_k$.*

*Proof.*

Let the $d$-dimensional output $\mathbf{z}'$ be given as

$$\begin{aligned}\mathbf{z}'^T &= \left(\mathbf{z}'_{U_0}\right)^T + \ldots + \left(\mathbf{z}'_{U_{K_{out}}}\right)^T \\ &= \mathbf{A}^\dagger_{U_0}\tilde{\mathbf{z}}_0 + \left(F^{r_0}_{U_0}(\Delta)\right)^T + \ldots + \mathbf{A}^\dagger_{U_{K_{out}-1}}\tilde{\mathbf{z}}_{K_{out}-1} + \left(F^{r_{K_{out}-1}}_{U_{K_{out}-1}}(\Delta)\right)^T,\end{aligned} \tag{139}$$

where $U_k$ is the $k$-th user, $k = 0, \ldots, K_{out} - 1$,

$$\tilde{\mathbf{z}}_k = diag\left(\tilde{z}_{U_k,0}, \ldots, \tilde{z}_{U_k,r_k-1}\right) \in \mathcal{CN}\left(0, \mathbf{K}_{\mathbf{z}_{U_k}}\right) \tag{140}$$

is the $r_k$-dimensional input of $U_k$ (referred to as codeword $\tilde{\mathbf{z}}_k$ of $U_k$), $r_k \leq d$, $\sum_{K_{out}} r_k = d$, and $\tilde{z}_{U_k,j} \in \mathcal{CN}\left(0, 2\sigma^2_{\omega_0}\right)$, and $\mathbf{A}_{U_k}$ is an $r_k$-dimensional vector with a $j$-th entry of $A_{U_k,j} = \frac{1}{l}\sum_{i=0}^{l-1} F_{U_i}\left(T_{j,i}(\mathcal{N}_{j,i})\right)$, that is, the averaged Fourier-transformed sub-channel coefficients of $U_k$, $j = 0, \ldots, r_k - 1$,

$$\begin{aligned}\mathbf{A}_{U_k} &= \left(A^\dagger_{U_k,0}, \ldots, A^\dagger_{U_k,r_k-1}\right)^T \\ &= \left[\tfrac{1}{l}\sum_{i=0}^{l-1} F_{U_k}\left(T_{0,i}(\mathcal{N}_{0,i})\right)^\dagger, \ldots, \tfrac{1}{l}\sum_{i=0}^{l-1} F_{U_k}\left(T_{r_k-1,i}(\mathcal{N}_{r_k-1,i})\right)^\dagger\right]^T,\end{aligned} \tag{141}$$

and

$$F^{r_k}_{U_k}(\Delta) = \left(F(\Delta_{U_k,0}), \ldots, F(\Delta_{U_k,r_k-1})\right)^T \in \mathcal{CN}\left(0, \mathbf{K}_{F^{r_k}_{U_k}(\Delta)}\right). \tag{142}$$

From $\mathbf{A}_{U_k}$, $\left|\mathbf{A}_{U_k}\right|$ is expressed as

$$\left|\mathbf{A}_{U_k}\right| = \sum_{w=0}^{r_k-1}\left|A_{U_k,w}\right| = \sum_{w=0}^{r_k-1}\left|\tfrac{1}{l}\sum_{i=0}^{l-1} F_{U_k}\left(T_{w,i}(\mathcal{N}_{w,i})\right)\right|, \tag{143}$$

where

$$\left|\mathbf{A}_{U_k}\right|\tilde{\mathbf{z}}_k = \sum_{j=0}^{r_k-1}\tfrac{1}{l}\sum_{i=0}^{l-1}\left|F\left(T_{j,i}(\mathcal{N}_{j,i})\right)\right|\tilde{\mathbf{z}}_k. \tag{144}$$

In the further parts of the proof, without loss of generality, we assume collective measurement for each $U_k$, and let the number input codewords be selected to $N = 2$, thus $\tilde{\mathbf{z}}_k \in \left\{\tilde{\mathbf{z}}_{A,k}, \tilde{\mathbf{z}}_{B,k}\right\}$. Then for any $U_k$, applying the results of Theorem 2 leads to

$$\begin{aligned}\nu^{r_k}_{U_k} &= \frac{\mathbf{A}^\dagger_{U_k}\tilde{\mathbf{z}}_{A,k} - \mathbf{A}^\dagger_{U_k}\tilde{\mathbf{z}}_{B,k}}{\left|\left(\mathbf{A}^\dagger_{U_k}\tilde{\mathbf{z}}_{A,k} - \mathbf{A}^\dagger_{U_k}\tilde{\mathbf{z}}_{B,k}\right)\right|} \\ &= \frac{\mathbf{A}^\dagger_{U_k}\left(\tilde{\mathbf{z}}_{A,k} - \tilde{\mathbf{z}}_{B,k}\right)}{\left|\mathbf{A}^\dagger_{U_k}\left(\tilde{\mathbf{z}}_{A,k} - \tilde{\mathbf{z}}_{B,k}\right)\right|} \\ &= \frac{\mathbf{A}^\dagger_{U_k}\mathbf{M}^2_{U_k}}{\left|\mathbf{A}^\dagger_{U_k}\mathbf{M}^2_{U_k}\right|},\end{aligned} \tag{145}$$

where $\mathbf{M}^2_{U_k}$ is the codeword difference matrix of $U_k$ at $N = 2$,

$$\mathbf{M}_{U_k}^2 = \tilde{\mathbf{z}}_{A,k} - \tilde{\mathbf{z}}_{B,k}. \tag{146}$$

In particular, projecting $\mathbf{z}'_{U_k}$ along $\nu_{U_k}^{r_k}$ via $\mathcal{P}(\mathbf{z}'_{U_k})$ yields a complex scalar $\Gamma_{U_k} \in \mathcal{C}$ as follows:

$$\begin{aligned} \mathcal{P}(\mathbf{z}'_{U_k}) &= \left(\nu_{U_k}^{r_k}\right)^{\dagger}\left[\left(\mathbf{z}'_{U_k}\right) - \tfrac{1}{2}\left(\mathbf{A}_{U_k}^{\dagger}\left(\tilde{\mathbf{z}}_{A,k} + \tilde{\mathbf{z}}_{B,k}\right)\right)^T\right] \\ &= s_{U_k}\left|\left(\mathbf{A}_{U_k}^{\dagger}\tilde{\mathbf{z}}_{A,k} - \mathbf{A}_{U_k}^{\dagger}\tilde{\mathbf{z}}_{B,k}\right)\right| + F_{U_k}(\Delta) \\ &= s_{U_k}\left|\mathbf{A}_{U_k}^{\dagger}\mathbf{M}_{U_k}^2\right| + F_{U_k}(\Delta) \\ &= \Gamma_{U_k}, \end{aligned} \tag{147}$$

where $s_{U_k} \in \{-0.5, 0.5\} \in \mathbb{R}$ and $F_{U_k}(\Delta) \in \mathcal{CN}(0, 2\sigma_\mathcal{N}^2)$ are scalars, which yields the $\mathcal{S}_{U_k}^{r_k}$ sufficient statistic of $U_k$ as

$$\begin{aligned} \mathcal{S}_{U_k}^{r_k} &= \left(\nu_{U_k}^{r_k}\right)^{\dagger}\mathbf{z}'_{U_k} \\ &= s_{U_k}\left|\mathbf{A}_{U_k}^{\dagger}\mathbf{M}_{U_k}^2\right| + F_{U_k}(\Delta). \end{aligned} \tag{148}$$

Note that a strictly suboptimal $\kappa(\mathbf{z}')$ decoding operation can also be defined by applying $\kappa$ on the $d$-dimensional $\mathbf{z}'$ in the phase space, such that

$$\begin{aligned} \kappa(\mathbf{z}') &= \mathbf{A}^{-1}\mathbf{z}' \\ &= \mathbf{A}^{-1}\left(\mathbf{A}\mathbf{z} + F^d(\Delta)\right) \\ &= \mathbf{z} + \mathbf{A}^{-1}F^d(\Delta), \end{aligned} \tag{149}$$

where $\mathbf{A}^{-1}$ is the inverse of $\mathbf{A}$,

$$\mathbf{A} = \left[\tfrac{1}{l}\sum_{i=0}^{l-1} F\left(T_{0,i}(\mathcal{N}_{0,i})\right)^{\dagger}, \ldots, \tfrac{1}{l}\sum_{i=0}^{l-1} F\left(T_{d-1,i}(\mathcal{N}_{d-1,i})\right)^{\dagger}\right]^T. \tag{150}$$

Precisely, in this case, the resulting $r_k$-dimensional scaled noise $\eta_{U_k}^{r_k}$,

$$\eta_{U_k}^{r_k} = \mathbf{A}^{-1}F^d(\Delta) \in \mathcal{CN}\left(0, \mathbf{K}_{\mu_{U_k}^{r_k}}\right), \tag{151}$$

of each $U_k$ user's is correlated; thus, (139) can be rewritten as

$$\begin{aligned} \mathbf{z}'^T &= \left(\mathbf{z}'_{U_0}\right)^T + \ldots + \left(\mathbf{z}'_{U_{K_{out}}}\right)^T \\ &= \mathbf{A}_{U_0}^{\dagger}\tilde{\mathbf{z}}_0 + \left(\eta_{U_0}^{r_0}\right)^T + \ldots + \mathbf{A}_{U_{K_{out}-1}}^{\dagger}\tilde{\mathbf{z}}_{K_{out}-1} + \left(\eta_{U_{K_{out}-1}}^{r_{K_{out}-1}}\right)^T, \end{aligned} \tag{152}$$

where $\eta_{U_k}$ is a $\mathcal{CN}\left(0, \sigma_{\eta_{U_k}}^2\right)$ distributed independent random variable.

In particular, applying a projection $\mathcal{P}(\kappa(\mathbf{z}'_{U_k}))$ on $\kappa(\mathbf{z}'_{U_k})$ leads to

$$\begin{aligned} \mathcal{P}(\kappa(\mathbf{z}'_{U_k})) &= \widehat{\mathbf{z}}'_{U_k} \\ &= \left(\partial_{U_k}^{\dagger}\mathbf{A}_{U_k}\right)\tilde{\mathbf{z}}_{U_k} + \Omega(\Delta), \end{aligned} \tag{153}$$

where $\Omega(\Delta) \in \mathcal{CN}(0,\sigma_\mathcal{N}^2)$ is a random variable (scaled noise), which is completely independent from the noise $F_{U_k}(\Delta)$ of $\mathbf{z}'_{U_k}$ of $U_k$, and $\partial_{U_k}$ is a $r_k$-dimensional vector expressed precisely as

$$\partial_{U_k} = \tfrac{1}{|\mathbf{A}_{U_k}|} \neg\left(\mathbf{A}^*_{U_k}\right), \tag{154}$$

where operator $\neg$ inverts the sign of some corresponding terms of $\mathbf{A}^*_{U_k}$ (* is the complex conjugate). Specifically, because the resulting $\Omega(\Delta)$ noise of each $U_k$ is independent from the noise obtained by $U_k$, the application of a $\kappa\left(\mathbf{z}'_{U_k}\right)$ operation on $\mathbf{z}'_{U_k}$ is strictly suboptimal, and $\mathcal{P}\left(\kappa\left(\mathbf{z}'_{U_k}\right)\right)$ results in a scaled scalar output $\widehat{z}'_{U_k}$.

The $(K_{in}, K_{out})$ multiuser setting defines a $K_{out}$-dimensional space, which is depicted in Fig. 5. Each user $U_k$ identifies a vector $\mathbf{A}_{U_k}$ and a sufficient statistics $\mathcal{S}^{r_k}_{U_k}$. The decoding of $\mathbf{z}'_{U_k}$ of $U_k$ is achieved via operation $\mathcal{P}\left(\mathbf{z}'_{U_k}\right)$.

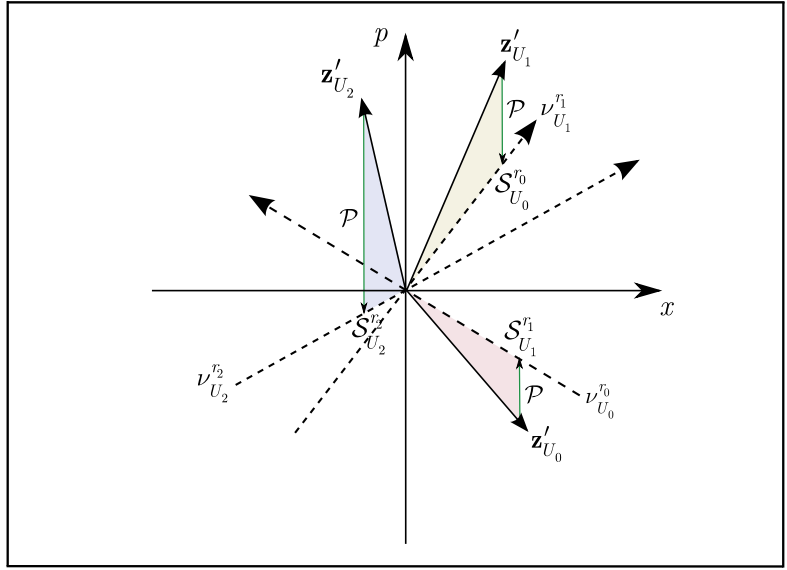

**Figure 5.** The $\mathcal{S}^{r_k}_{U_k}$ sufficient statics in the $K_{out}$-dimensional space for a multiuser multicarrier CVKQD decoding, for $K_{out} = 3$.

To derive the error probability, we can directly apply the results of Theorem 2. Let $N = 2$; thus, in a $(K_{in}, K_{out})$ multiuser setting, for a particular user $U_k$, (135) leads to an error probability precisely,

$$\Pr\left(\tilde{\mathbf{z}}_{A,k} \to \tilde{\mathbf{z}}_{B,k}\right) \leq \mathbb{E}\left[\prod_{j=0}^{r_k-1} \frac{1}{1+\mathrm{SNR}_{U_k}\frac{1}{4}\lambda_j^2}\right], \tag{155}$$

where $\lambda_j^2$ is the $j$-th singular value of $\mathbf{M}^2_{U_k} = \tilde{\mathbf{z}}_{A,k} - \tilde{\mathbf{z}}_{B,k}$, and $\mathrm{SNR}_{U_k}$ is the SNR of $U_k$. The results can be extended for arbitrary $N$ and combination of $(K_{in}, K_{out})$.

∎

# 6 Conclusions

The multicarrier CVQKD transmission utilizes subcarrier CVs for information transmission. We introduced the adaptive quadrature detection for multicarrier CVQKD. The proposed decoding scheme exploits the statistics of the sub-channels, which are provided by our sub-channel estimation phase. The sub-channel estimation procedure scans through the conditions of the sub-channels via pilot-subcarrier CVs. The error probability of the sub-channel estimation reaches the theoretical minimum via our subcarrier spreading technique. Adaptive quadrature decoding can be performed with homodyne or heterodyne measurement, single or collective measurement setting. We also extended the adaptive quadrature detection for a multiuser multicarrier CVQKD setting. The adaptive quadrature detection scheme provides a flexible framework to extract the potential of multicarrier CVQKD, specifically functional and convenient for experimental scenarios.

# Acknowledgements

The author would like to thank Professor Sandor Imre for useful discussions. This work was partially supported by the GOP-1.1.1-11-2012-0092 (*Secure quantum key distribution between two units on optical fiber network*) project sponsored by the EU and European Structural Fund, and by the COST Action MP1006.

# Supplemental Information

## S.1 Notations

The notations of the manuscript are summarized in Table S.1.

**Table S.1.** Summary of notations.

| Symbol | Description |
|---|---|
| $i$ | Index for the $i$-th subcarrier Gaussian CV, $\left|\phi_i\right\rangle = x_i + \mathrm{i}p_i$. |
| $j$ | Index for the $j$-th Gaussian single-carrier CV, $\left|\varphi_j\right\rangle = x_j + \mathrm{i}p_j$. |
| $l$ | Number of Gaussian sub-channels $\mathcal{N}_i$ for the transmission of the Gaussian subcarriers. The overall number of the sub-channels is $n$. The remaining $n-l$ sub-channels do not transmit valuable information. |
| $x_i, p_i$ | Position and momentum quadratures of the $i$-th Gaussian subcarrier, $\left|\phi_i\right\rangle = x_i + \mathrm{i}p_i$. |
| $x'_i, p'_i$ | Noisy position and momentum quadratures of Bob's $i$-th noisy subcarrier Gaussian CV, $\left|\phi'_i\right\rangle = x'_i + \mathrm{i}p'_i$. |
| $x_j, p_j$ | Position and momentum quadratures of the $j$-th Gaussian single-carrier $\left|\varphi_j\right\rangle = x_j + \mathrm{i}p_j$. |
| $x'_j, p'_j$ | Noisy position and momentum quadratures of Bob's $j$-th recovered single-carrier Gaussian CV $\left|\varphi'_j\right\rangle = x'_j + \mathrm{i}p'_j$. |
| $x_{A,i}, p_{A,i}$ | Alice's quadratures in the transmission of the $i$-th subcarrier. |
| $\left|\phi_i\right\rangle, \left|\phi'_i\right\rangle$ | Transmitted and received Gaussian subcarriers. The subcarriers have angles $\theta_i^* \in [0, 2\pi]$, $\theta_i \in [0, 2\pi]$ CVs in the phase space $\mathcal{S}$. |
| $Q(\cdot)$ | Gaussian tail function. |
| $p_i$ | Pilot-subcarrier CV for the $\mathcal{S}_i(\mathcal{N}_i) \in \mathcal{C}$ channel statistics |
| | of sub-channel $\mathcal{N}_i$. Complex variable, $p_i = \mathrm{Re}(p_i) + \mathrm{Im}(p_i) = x_{p_i} + \mathrm{i}p_{p_i} \in \mathcal{C}$, $|p_i| > 0$, where $x_{p_i}, p_{p_i}$ identify the position and momentum quadratures. Used for the estimation of $F(T_i(\mathcal{N}_i))$ of $\mathcal{N}_i$. Also expressed as $p_i = F^{-1}(p_j)$, where $F^{-1}$ is the inverse Fourier operation. |
| $p_j$ | Pilot single-carrier CV for the $\mathcal{N}_j$ single-carrier channel estimation. Complex variable, $p_j = F(p_i) = x_{p_j} + \mathrm{i}p_{p_j}$, where $x_{p_j}, p_{p_j}$ identify the position and momentum quadratures. |
| $\mathcal{S}_i(\mathcal{N}_i)$ | Sufficient statistic for the estimation of $F(T_i(\mathcal{N}_i))$ of sub-channel $\mathcal{N}_i$, complex variable. Evaluated as $\mathcal{S}_i(\mathcal{N}_i) = \varsigma_i^\dagger p_i' = \left(p_i^\dagger / |p_i|^2\right) p_i' = F(T_i(\mathcal{N}_i)) + F'(\Delta)$. |
| $\mathcal{S}(\mathcal{N}_j)$ | Sufficient statistic for the estimation of $\frac{1}{l}\sum_l F(T_i(\mathcal{N}_i))$ of the single-carrier channel $\mathcal{N}_j$, expressed as $$\mathcal{S}(\mathcal{N}_j) = \varsigma_j^\dagger \mathbf{q}_j' = A_j + F'(\Delta),$$ where $\varsigma_j = \mathbf{q}_j / |\mathbf{q}_j|^2$, $\mathbf{q}_j = (p_0, ..., p_{l-1})^T \in \mathcal{C}^l$, $\mathbf{q}_j' = (p_0', ..., p_{l-1}')^T \in \mathcal{C}^l$, $A_j = \frac{1}{l}\left(\sum_{i=0}^{l-1} F(T_{j,i}(\mathcal{N}_{j,i}))\right) \in \mathcal{C}$, $F'(\Delta) \in \mathcal{CN}\left(0, 2\sigma_\mathcal{N}^2 / |\mathbf{q}_j|^2\right)$. |
| $\mathcal{P}(\cdot)$ | Projector. |
| $\Gamma$ | The result of projection $\mathcal{P}(\cdot)$, identifies the $\mathcal{S}$ sufficient statistics in the $\mathcal{C}$ complex scalar space. |
| $p_i'$ | Noisy pilot-subcarrier CV, complex variable, $p_i' = F(T_i(\mathcal{N}_i)) p_i + F(\Delta)$. |
| $\varsigma_i$ | Complex variable, $\varsigma_i = p_i / |p_i|^2$ to derive the $\mathcal{S}(\mathcal{N}_i)$ sufficient sub-channel channel statics of $\mathcal{N}_i$. |

| | |
|---|---|
| $s$ | Scalar real variable, $s \in \{-0.5, 0.5\} \in \mathbb{R}$. |
| $\gamma$ | Subspace of one $\mathbb{R}$ real dimension. |
| $\mathcal{E}$ | Mean square error. |
| $\zeta(\cdot)$ | Linear estimation operator. |
| $E$ | Minimum mean squared error. |
| $C \in \mathcal{C}$ | Constant complex variable to derive $\zeta(\cdot)$. |
| $\mathbf{C} \in \mathcal{C}^l$ | Constant complex vector in $\mathcal{C}^l$ to derive $\zeta(\cdot)$ of the single-carrier channel $\mathcal{N}_j$. |
| $\mathbf{q}_j$ | An $l$-dimensional complex pilot-subcarrier CV vector, conveys the $l$ pilot subcarrier CVs, $\mathbf{q}_j = \left(p_{j,0}, \ldots, p_{j,l-1}\right)^T \in \mathcal{C}^l$. The noisy vector is $\mathbf{q}'_j = \left(p'_{j,0}, \ldots, p'_{j,l-1}\right)^T \in \mathcal{C}^l$. |
| $\mathbf{q}_x \in \mathcal{C}^g$ | A $g$-dimensional complex pilot-subcarrier CV vector $\mathbf{q}_x = \left(p_{x,0}, \ldots, p_{x,g-1}\right)^T$, defined for the subcarrier spreading, $g + (l-1) = n$, where $n$ is the overall number of the sub-channels, $l$ is the number of sub-channels that are utilized for transmission in the private transmission phase. |
| $\mathbf{p}_x$ | An $l$-dimensional complex pilot-subcarrier CV vector, $\mathbf{p}_x \in \mathcal{C}^l$, $\mathbf{p}_x = \left(p_{x,0}, \ldots, p_{x,l-1}\right)^T$, where $p_{x,i}$ is the constant pilot-subcarrier CV of sub-channel $\mathcal{N}_i, i = 0, \ldots, l-1$, such that, $p_{x,i} = p_x \in \mathcal{C}, i = 0, \ldots, l-1$. |
| $\mathbf{p}'_x$ | The noisy version of $\mathbf{p}_x$, $\mathbf{p}'_x = \left(p'_{x,0}, \ldots, p'_{x,l-1}\right)^T$, $l$-dimensional constant complex pilot-subcarrier CV vector. |
| $\mathbf{P}^i_x = (\mathbf{a}_0, \mathbf{q}_x, \mathbf{b}_0)$ | An $n$-dimensional complex vector $\mathbf{P}^i_x \in \mathcal{C}^n$, defined for the subcarrier spreading technique, $i = 0, \ldots, l-1$, $\mathbf{q}_x = \left(p_{x,0}, \ldots, p_{x,g-1}\right)^T$, $\mathbf{a}_0$ is an $i$-dimensional vector, $\mathbf{a}_0 = \left(\rho^0_0, \ldots, \rho^0_{i-1}\right)^T$, where $\rho^0_w = |0\rangle\langle 0|, w = 0, \ldots, i-1$, $\mathbf{b}_0$ is an $u = (l-1) - i$-dimensional vector, $\mathbf{b}_0 = \left(\rho^0_0, \ldots, \rho^0_{u-1}\right)^T$, where $\rho^0_w = |0\rangle\langle 0|, w = 0, \ldots, u-1$. |
| $\mathbf{V}^i(\Delta)$ | An $n$-dimensional noise vector, defined for the subcarrier |
| | spreading technique. |
| $\varsigma_x = \mathbf{P}_x^i \big/ \big|\mathbf{P}_x^i\big|$ | Sufficient statistics for the estimation of sub-channel $\mathcal{N}_i$ in subcarrier spreading. Complex vector, $\varsigma_x = \mathbf{P}_x^i \big/ \big|\mathbf{P}_x^i\big|$. |
| $\Upsilon$ | Real variable, $\Upsilon = \frac{|\mathbf{p}_x|^2}{g} = |p_x|^2$, where $\mathbf{p}_x = (p_{x,0},...,p_{x,g-1})^T$. |
| $\mathbf{M}^N$ | Codeword difference matrix of the $N$ codewords $\{\mathbf{z}_0,...,\mathbf{z}_{N-1}\}$. Hermitian matrix, $(\mathbf{M}^N)^\dagger = \mathbf{M}^N$. |
| $\lambda_i^2$ | The $i$-th singular value of $\mathbf{M}^N$, $\tau = diag(\lambda_0^2,...,\lambda_{d-1}^2)$. |
| $\mathbf{A}$ | A $d$-dimensional vector of the averaged transmittance coefficients, evaluated as $\mathbf{A} = \left(\frac{1}{l}\sum_{i=0}^{l-1} F\left(T_{0,i}\left(\mathcal{N}_{0,i}\right)\right)^\dagger,...,\frac{1}{l}\sum_{i=0}^{l-1} F\left(T_{d-1,i}\left(\mathcal{N}_{d-1,i}\right)\right)^\dagger\right)^T$. |
| $\mathbf{z} \in \mathcal{CN}(0,\mathbf{K}_\mathbf{z})$ | A $d$-dimensional input CV vector to transmit valuable information. |
| $\mathbf{z}'^T$ | A $d$-dimensional noisy output vector, $\mathbf{z}'^T = \mathbf{A}^\dagger \mathbf{z} + \left(F^d(\Delta)\right)^T = (z'_0,...,z'_{d-1})$, where $z'_j = \left(\frac{1}{l}\sum_{i=0}^{l-1} F\left(T_{j,i}\left(\mathcal{N}_{j,i}\right)\right)\right)z_j + F(\Delta) \in \mathcal{CN}\left(0,2\left(\sigma^2_{\omega_0} + \sigma^2_\mathcal{N}\right)\right)$. |
| $\mathbf{G}_{uv}$ | An orthogonal matrix, $uv$ indexes the row and column. |
| $\mathcal{H}_{Az_A}, \mathcal{H}_{Az_B}$ | Hyperplanes of the decision region, while $Az_A, Az_B \in \mathcal{C}$ and $z'_A, z'_B \in \mathcal{C}$ are complex scalar quantities. |
| $\mu_{Az_A}, \mu_{Az_B} \in \mathcal{C}$ | Expected means of $\Gamma$, complex variables. |
| $U_k$ | Identifies user $U_k$ in a multiuser scenario. |
| $(K_{in}, K_{out})$ | Multiuser setting, number of transmitter and receiver users. |
| $\mathbf{z}'^T$ | A $d$-dimensional output in a multiuser setting, evaluated as $\left(\mathbf{z}'_{U_0}\right)^T + ... + \left(\mathbf{z}'_{U_{K_{out}}}\right)^T$ $= \mathbf{A}^\dagger_{U_0} \tilde{\mathbf{z}}_0 + \left(F^{r_0}_{U_0}(\Delta)\right)^T + ... + \mathbf{A}^\dagger_{U_{K_{out}-1}} \tilde{\mathbf{z}}_{K_{out}-1} + \left(F^{r_{K_{out}-1}}_{U_{K_{out}-1}}(\Delta)\right)^T$. |
| $\tilde{\mathbf{z}}_k$ | The $r_k$-dimensional codeword of user $U_k$, diagonal matrix, $r_k \leq d$, $\sum_{K_{out}} r_k = d$, |
| | $$\tilde{\mathbf{z}}_k = diag\left(\tilde{z}_{U_k,0},...,\tilde{z}_{U_k,r_k-1}\right) \in \mathcal{CN}\left(0,\mathbf{K}_{\mathbf{z}_{U_k}}\right),$$ where $\tilde{z}_{U_k,j} \in \mathcal{CN}\left(0,2\sigma^2_{\omega_0}\right)$. |
| $\mathbf{A}_{U_k}$ | An $r_k$-dimensional vector that contains the $\frac{1}{l}\sum_{i=0}^{l-1} F_{U_i}\left(T_{j,i}\left(\mathcal{N}_{j,i}\right)\right)$ averaged $F_{U_i}$ Fourier transformed sub-channel coefficients of $\tilde{z}_j$ for $U_k$, $j=0,...,r_k-1$. |
| $\kappa(\cdot)$ | A suboptimal operation for the pre-decoding $d$-dimensional $\mathbf{z}'$ in the phase space. |
| $\eta_{U_k}^{r_k}$ | The resulting $r_k$-dimensional scaled noise vector $\eta_{U_k}^{r_k}$, $\eta_{U_k}^{r_k} = \mathbf{A}^{-1} F^d(\Delta) \in \mathcal{CN}\left(0,\mathbf{K}_{\mu_{U_k}^{r_k}}\right)$. The $\eta_{U_k}$ projected noise is a $\mathcal{CN}\left(0,\sigma^2_{\eta_{U_k}}\right)$ distributed independent random variable, evaluated as $$\begin{aligned}\kappa(\mathbf{z}') &= \mathbf{A}^{-1}\mathbf{z}'\\ &= \mathbf{A}^{-1}\left(\mathbf{A}\mathbf{z}+F^d(\Delta)\right)\\ &= \mathbf{z}+\mathbf{A}^{-1}F^d(\Delta).\end{aligned}$$ |
| $\partial_{U_k}$ | An $r_k$-dimensional vector of user $U_k$, expressed $\partial_{U_k} = \frac{1}{|\mathbf{A}_{U_k}|}\neg\left(\mathbf{A}^*_{U_k}\right)$, where operator $\neg$ inverts the sign of some corresponding terms of $\mathbf{A}^*_{U_k}$, where $*$ is the complex conjugate. |
| $\Omega(\Delta) \in \mathcal{CN}\left(0,\sigma^2_{\mathcal{N}}\right)$ | Random variable (scaled noise), independent from the noise $F_{U_k}(\Delta)$ of $\mathbf{z}'_{U_k}$ of $U_k$, derived as $$\begin{aligned}\mathcal{P}\left(\kappa\left(\mathbf{z}'_{U_k}\right)\right) &= \hat{z}'_{U_k}\\ &= \left(\partial^{\dagger}_{U_k}\mathbf{A}_{U_k}\right)\tilde{z}_{U_k} + \Omega(\Delta).\end{aligned}$$ |
| $\widehat{\mathrm{SNR}}$ | A scaled SNR quantity, $\widehat{\mathrm{SNR}} = \frac{0.5|p_x|^2}{2\sigma^2_{\mathcal{N}}} = \frac{1}{2}\mathrm{SNR}$. |
| $\mathrm{SNR}_{U_k}$ | The SNR quantity of $U_k$. |
| $p_{err}\left(p'_x\big|\zeta\left(F\left(\mathbf{v}(\mathcal{N})\right)\right)\right)$ | Conditional error probability of decoding pilot CV $p'_x$ using the linear estimate $\zeta\left(F\left(\mathbf{v}(\mathcal{N})\right)\right)$. |
| E | An error event, defined as $\mathrm{E}: \left|\zeta\left(F\left(\mathbf{v}(\mathcal{N})\right)\right)\right|^2 < \frac{1}{\widehat{\mathrm{SNR}}}$. |

| | |
|---|---|
| $A_j$ | Complex variable, $A_j = \frac{1}{l}\left(\sum_{i=0}^{l-1} F\left(T_{j,i}\left(\mathcal{N}_{j,i}\right)\right)\right) \in \mathcal{C}$, the averaged transmittance coefficient of the single-carrier channel $\mathcal{N}_j$. |
| $\chi(\cdot)$ | Function that separately identifies the $x$ position or $p$ momentum quadrature components as $\text{Re}(\cdot)$ and $\text{Im}(\cdot)$, respectively. |
| $\vartheta$ | Variable to identify the noise of the estimation process on the $x$ position and $p$ momentum quadrature components, $\vartheta = \frac{\left(\chi(\mathbf{C})^T \chi(\mathbf{q}_j)\right)^2 \mathbb{E}\left[\chi(A_j)^2\right]}{\left|\chi(\mathbf{C})\right|^2 \sigma_{\mathcal{N}}^2}$. |
| $\left\{\mathcal{C}_{\mathbf{z}_A}, \mathcal{C}_{\mathbf{z}_B}\right\}, \left\{\mathcal{C}_{\mathbf{p}_A}, \mathcal{C}_{\mathbf{p}_B}\right\}$ | Phase-space constellations of $\mathbf{z}_A, \mathbf{z}_B$, and private codewords $\mathbf{p}_A, \mathbf{p}_B$. |
| $U_{K_{out}}$ | The unitary CVQFT operation, $U_{K_{out}} = \frac{1}{\sqrt{K_{out}}} e^{\frac{-i2\pi ik}{K_{out}}}$, $i, k = 0, \ldots, K_{out} - 1$, $K_{out} \times K_{out}$ unitary matrix. |
| $U_{K_{in}}$ | The unitary inverse CVQFT operation, $U_{K_{in}} = \frac{1}{\sqrt{K_{in}}} e^{\frac{i2\pi ik}{K_{in}}}$, $i, k = 0, \ldots, K_{in} - 1$, $K_{in} \times K_{in}$ unitary matrix. |
| $z \in \mathcal{CN}\left(0, \sigma_z^2\right)$ | The variable of a single-carrier Gaussian CV state, $\left|\varphi_i\right\rangle \in \mathcal{S}$. Zero-mean, circular symmetric complex Gaussian random variable, $\sigma_z^2 = \mathbb{E}\left[\left|z\right|^2\right] = 2\sigma_{\omega_0}^2$, with i.i.d. zero mean, Gaussian random quadrature components $x, p \in \mathbb{N}\left(0, \sigma_{\omega_0}^2\right)$, where $\sigma_{\omega_0}^2$ is the variance. |
| $\Delta \in \mathcal{CN}\left(0, \sigma_\Delta^2\right)$ | The noise variable of the Gaussian channel $\mathcal{N}$, with i.i.d. zero-mean, Gaussian random noise components on the position and momentum quadratures $\Delta_x, \Delta_p \in \mathbb{N}\left(0, \sigma_{\mathcal{N}}^2\right)$, $\sigma_\Delta^2 = \mathbb{E}\left[\left|\Delta\right|^2\right] = 2\sigma_{\mathcal{N}}^2$. |
| $d \in \mathcal{CN}\left(0, \sigma_d^2\right)$ | The variable of a Gaussian subcarrier CV state, $\left|\phi_i\right\rangle \in \mathcal{S}$. Zero-mean, circular symmetric Gaussian random variable, $\sigma_d^2 = \mathbb{E}\left[\left|d\right|^2\right] = 2\sigma_\omega^2$, with i.i.d. zero mean, Gaussian ran- |

| | |
|---|---|
| | dom quadrature components $x_d, p_d \in \mathbb{N}\left(0, \sigma_\omega^2\right)$, where $\sigma_\omega^2$ is the modulation variance of the Gaussian subcarrier CV state. |
| $F^{-1}(\cdot) = \text{CVQFT}^\dagger(\cdot)$ | The inverse CVQFT transformation, applied by the encoder, continuous-variable unitary operation. |
| $F(\cdot) = \text{CVQFT}(\cdot)$ | The CVQFT transformation, applied by the decoder, continuous-variable unitary operation. |
| $F^{-1}(\cdot) = \text{IFFT}(\cdot)$ | Inverse FFT transform, applied by the encoder. |
| $\sigma_{\omega_0}^2$ | Single-carrier modulation variance. |
| $\sigma_\omega^2 = \tfrac{1}{l}\sum_l \sigma_{\omega_i}^2$ | Multicarrier modulation variance. Average modulation variance of the $l$ Gaussian sub-channels $\mathcal{N}_i$. |
| $\begin{aligned}\left|\phi_i\right\rangle &= \left|\text{IFFT}\left(z_{k,i}\right)\right\rangle \\ &= \left|F^{-1}\left(z_{k,i}\right)\right\rangle = \left|d_i\right\rangle.\end{aligned}$ | The $i$-th Gaussian subcarrier CV of user $U_k$, where IFFT stands for the Inverse Fast Fourier Transform, $\left|\phi_i\right\rangle \in \mathcal{S}$, $d_i \in \mathcal{CN}\left(0, \sigma_{d_i}^2\right)$, $\sigma_{d_i}^2 = \mathbb{E}\left[\left|d_i\right|^2\right]$, $d_i = x_{d_i} + \mathrm{i} p_{d_i}$, $x_{d_i} \in \mathbb{N}\left(0, \sigma_{\omega_F}^2\right)$, $p_{d_i} \in \mathbb{N}\left(0, \sigma_{\omega_F}^2\right)$ are i.i.d. zero-mean Gaussian random quadrature components, and $\sigma_{\omega_F}^2$ is the variance of the Fourier transformed Gaussian state. |
| $\left|\varphi_{k,i}\right\rangle = \text{CVQFT}\left(\left|\phi_i\right\rangle\right)$ | The decoded single-carrier CV of user $U_k$ from the subcarrier CV, expressed as $F\left(\left|d_i\right\rangle\right) = \left|F\left(F^{-1}\left(z_{k,i}\right)\right)\right\rangle = \left|z_{k,i}\right\rangle$. |
| $\mathcal{N}$ | Gaussian quantum channel. |
| $\mathcal{N}_i, i = 0,\ldots,n-1$ | Gaussian sub-channels. |
| $T(\mathcal{N})$ | Channel transmittance, normalized complex random variable, $T(\mathcal{N}) = \operatorname{Re} T(\mathcal{N}) + \mathrm{i}\operatorname{Im} T(\mathcal{N}) \in \mathcal{C}$. The real part identifies the position quadrature transmission, the imaginary part identifies the transmittance of the position quadrature. |
| $T_i(\mathcal{N}_i)$ | Transmittance coefficient of Gaussian sub-channel $\mathcal{N}_i$, $T_i(\mathcal{N}_i) = \operatorname{Re}\left(T_i(\mathcal{N}_i)\right) + \mathrm{i}\operatorname{Im}\left(T_i(\mathcal{N}_i)\right) \in \mathcal{C}$, quantifies the position and momentum quadrature transmission, with (normalized) real and imaginary parts |
| | $0 \leq \operatorname{Re} T_i(\mathcal{N}_i) \leq 1/\sqrt{2}$, $\quad 0 \leq \operatorname{Im} T_i(\mathcal{N}_i) \leq 1/\sqrt{2}$, where $\operatorname{Re} T_i(\mathcal{N}_i) = \operatorname{Im} T_i(\mathcal{N}_i)$. |
| $T_{Eve}$ | Eve's transmittance, $T_{Eve} = 1 - T(\mathcal{N})$. |
| $T_{Eve,i}$ | Eve's transmittance for the $i$-th subcarrier CV. |
| $\mathbf{z} = \mathbf{x} + i\mathbf{p} = (z_0,...,z_{d-1})^T$ | A $d$-dimensional, zero-mean, circular symmetric complex random Gaussian vector that models $d$ Gaussian CV input states, $\mathcal{CN}(0, \mathbf{K_z})$, $\mathbf{K_z} = \mathbb{E}[\mathbf{zz}^\dagger]$, where $z_i = x_i + ip_i$, $\mathbf{x} = (x_0,...,x_{d-1})^T$, $\mathbf{p} = (p_0,...,p_{d-1})^T$, $x_i \in \mathbb{N}(0, \sigma^2_{\omega_0})$, $p_i \in \mathbb{N}(0, \sigma^2_{\omega_0})$ i.i.d. zero-mean Gaussian random variables. |
| $\mathbf{d} = F^{-1}(\mathbf{z})$ | An $l$-dimensional, zero-mean, circular symmetric complex random Gaussian vector, $\mathcal{CN}(0, \mathbf{K_d})$, $\mathbf{K_d} = \mathbb{E}[\mathbf{dd}^\dagger]$, $\mathbf{d} = (d_0,...,d_{l-1})^T$, $d_i = x_i + ip_i$, $x_i, p_i \in \mathbb{N}(0, \sigma^2_{\omega_F})$ are i.i.d. zero-mean Gaussian random variables, $\sigma^2_{\omega_F} = 1/\sigma^2_{\omega_0}$. The $i$-th component is $d_i \in \mathcal{CN}(0, \sigma^2_{d_i})$, $\sigma^2_{d_i} = \mathbb{E}[|d_i|^2]$. |
| $\mathbf{y}_k \in \mathcal{CN}(0, \mathbb{E}[\mathbf{y}_k \mathbf{y}_k^\dagger])$ | A $d$-dimensional zero-mean, circular symmetric complex Gaussian random vector. |
| $y_{k,m}$ | The $m$-th element of the $k$-th user's vector $\mathbf{y}_k$, expressed as $y_{k,m} = \sum_l F(T_i(\mathcal{N}_i))F(d_i) + F(\Delta_i)$. |
| $F(\mathbf{T}(\mathcal{N}))$ | Fourier transform of $\mathbf{T}(\mathcal{N}) = [T_0(\mathcal{N}_0)...T_{l-1}(\mathcal{N}_{l-1})]^T \in \mathcal{C}^l$, the complex transmittance vector. |
| $F(\Delta)$ | Complex vector, expressed as $F(\Delta) = e^{\frac{-F(\Delta)^T \mathbf{K}_{F(\Delta)} F(\Delta)}{2}}$, with covariance matrix $\mathbf{K}_{F(\Delta)} = \mathbb{E}[F(\Delta)F(\Delta)^\dagger]$. |
| $\mathbf{y}[j]$ | AMQD block, $\mathbf{y}[j] = F(\mathbf{T}(\mathcal{N}))F(\mathbf{d})[j] + F(\Delta)[j]$. |
| $\tau = \|F(\mathbf{d})[j]\|^2$ | An exponentially distributed variable, with density $f(\tau) = (1/2\sigma^{2n}_\omega)e^{-\tau/2\sigma^2_\omega}$, $\mathbb{E}[\tau] \leq n2\sigma^2_\omega$. |
| $T_{Eve,i}$ | Eve's transmittance on the Gaussian sub-channel $\mathcal{N}_i$, |
| | $T_{Eve,i} = \operatorname{Re} T_{Eve,i} + \mathrm{i}\operatorname{Im} T_{Eve,i} \in \mathcal{C}\,, \quad 0 \leq \operatorname{Re} T_{Eve,i} \leq 1/\sqrt{2}\,,$ $0 \leq \operatorname{Im} T_{Eve,i} \leq 1/\sqrt{2}\,,\ 0 \leq \left|T_{Eve,i}\right|^2 < 1\,.$ |
| $d_i$ | A $d_i$ subcarrier in an AMQD block. |
| $\nu_{\min}$ | The $\min\{\nu_0,\ldots,\nu_{l-1}\}$ minimum of the $\nu_i$ sub-channel coefficients, where $\nu_i = \sigma_{\mathcal{N}}^2 \big/ \left|F\left(T_i\left(\mathcal{N}_i\right)\right)\right|^2$ and $\nu_i < \nu_{Eve}$. |
| $\sigma_\omega^2$ | Modulation variance, $\sigma_\omega^2 = \nu_{Eve} - \nu_{\min}\mathcal{G}(\delta)_{p(x)}$, where $\nu_{Eve} = \tfrac{1}{\lambda},\ \lambda = \left|F\left(T_{\mathcal{N}}^*\right)\right|^2 = \tfrac{1}{n}\sum_{i=0}^{n-1}\left|\sum_{k=0}^{n-1} T_k^* e^{\tfrac{-\mathrm{i}2\pi ik}{n}}\right|^2$ and $T_{\mathcal{N}}^*$ is the expected transmittance of the Gaussian sub-channels under an optimal Gaussian collective attack. |

## S.2 Abbreviations

| | |
|---|---|
| **AMQD** | **Adaptive Multicarrier Quadrature Division** |
| **CV** | **Continuous-Variable** |
| **CVQFT** | **Continuous-Variable Quantum Fourier Transform** |
| **CVQKD** | **Continuous-Variable Quantum Key Distribution** |
| **DV** | **Discrete Variable** |
| **FFT** | **Fast Fourier Transform** |
| **IFFT** | **Inverse Fast Fourier Transform** |
| **MQA** | **Multiuser Quadrature Allocation** |
| **QKD** | **Quantum Key Distribution** |
| **SNR** | **Signal to Noise Ratio** |